\documentclass[twocolumn,showpacs,preprintnumbers,amsmath,amssymb]{revtex4-1}

\usepackage{graphicx}
\usepackage{dcolumn}
\usepackage{bm}

\newcommand{\be}{\begin{eqnarray}}
\newcommand{\ee}{\end{eqnarray}}
\newcommand{\nn}{\nonumber\\}
\newcommand{\nin}{\noindent}
\newcommand{\la}{\langle}
\newcommand{\ra}{\rangle}

\renewcommand{\theequation}{\arabic{section}.\arabic{equation}}

\begin{document}

\title{Antiferromagnetic Order and Bose-Einstein Condensation in 
Strongly-Correlated Cold-Atom Systems:
Bosonic $t$-$J$ Model in the Double-CP$^1$ Representation}

\date{\today}

\author{Yuki Nakano$^1$} 
\author{Takumi Ishima$^2$} 
\author{Naohiro Kobayashi$^2$} 
\author{Kazuhiko Sakakibara$^3$}
\author{Ikuo Ichinose$^2$}
\author{Tetsuo Matsui$^1$}
\affiliation{%
${}^1$ Department of Physics, Kinki University, 
Higashi-Osaka, 577-8502 Japan
}%
 \affiliation{${}^2$ Department of Applied Physics,
Nagoya Institute of Technology, 
Nagoya, 466-8555 Japan 
}
\affiliation{%
${}^3$ Department of Physics, Nara National College of Technology, 
Yamatokohriyama, 639-1080 Japan
}%


\begin{abstract}
We study the three-dimensional bosonic $t$-$J$  model, i.e.,
the $t$-$J$ model of ``bosonic electrons" at finite temperatures.
This model describes a system of cold bosonic atoms with two species
in an optical lattice. The model is derived from the Hubbard model
for  very large on-site repulsive interaction between
bosons of same species (hard-core nature) 
and also strong correlations between
different species.
The operator $B_{x\sigma}$ for an atom at the site $x$ 
with a two-component (pseudo-) spin $\sigma (=1,2)$ 
is treated as  a hard-core boson operator, 
and represented by
a composite of two slave particles;
a spinon described by a CP$^1$ field (Schwinger boson) $z_{x\sigma}$
and a holon described by a hard-core-boson field $\phi_x$
as $B_{x\sigma}=\phi^\dag_x z_{x\sigma}$. 
$\phi_x$ is then expressed
by a pseudo-spin, which is, in turn, represented by
another CP$^1$ (pseudo) spinon $w_{x\eta}$ as 
$\phi_x = w_{x2}^\dag w_{x1}$. 
We then have a double-CP$^1$ representation of the model
by $z_{x\sigma}$ and $w_{x\eta}$.
By means of Monte Carlo simulations of this bosonic $t$-$J$ model, 
we study its phase structure
and the possible  phenomena like appearance of antiferromagnetic 
long-range order, Bose-Einstein condensation, phase separation, etc.
They should be compared with the possible experimental
results of a recently studied boson-boson mixture  
like ${}^{87}$Rb and ${}^{41}$K in an optical lattice.

\end{abstract}

\maketitle

\section{Introduction}
\setcounter{equation}{0} 

Cold atoms\cite{cold_boson, cold} trapped in optical lattices
are one of the most interesting systems 
in condensed matter physics.
They are flexible because one can experimentally adjust their characteristics 
such as statistics and density of atoms, dimensionality of the system,  
signature and strength of interactions, etc.

Not only systems of single kind of atoms, but also systems
of atoms of two species like a mixture of bosons and fermions 
are studied experimentally\cite{fermibose}.
Recently, a boson-boson system of ${}^{87}$Rb and ${}^{41}$K
in a three-dimensional (3D) optical lattice 
has been  studied to produce  Bose-Einstein condensate\cite{bosonboson}.
The interactions among these atoms are also controllable\cite{bosonboson2}.

A standard model of cold {\it bosonic} atoms with two species 
may be the bosonic Hubbard model.
The two species of bosons may be described 
by a (pseudo-)spin $s=1/2$ degrees of freedom.
For very large strong repulsive interactions between atoms
 of {\it same species}, one may treat each boson as a hard-core boson (HCB).
Thus, usual electrons of Hubbard model 
are to be replaced here by HCB with spins.

From this HCB Hubbard model, one can derive 
the {\it bosonic} $t$-$J$ model as its low-energy effective model
for a large on-site repulsion between the opposite spins
(different species)
and small hole concentrations\cite{derivation, derivation2}.

 By using this bosonic $t$-$J$ model, 
 Boninsegni and Prokof'ev\cite{btj}
studied the interplay of magnetic ordering of pseudo-spins 
and Bose-Einstein condensation (BEC)/superfluidity (SF) 
of bosonic atoms.
By quantum Monte Carlo (MC) simulations, they
studied the low-temperature ($T$) phase diagram
of the two-dimensional (2D) model
for the case of 
{\it anisotropic spin coupling} $J_{x,y}=\alpha J_z, \alpha < 1$ and
$J_z\equiv J < t$,
and found the coexistence region of antiferromagnetic (AF) order and SF
as a result of the {\it phase separation} (PS) of hole-free (AF) and 
hole-rich (SF) phases.

In the previous paper\cite{aoki} we studied
the  bosonic $t$-$J$ model with the isotropic coupling
($\alpha=1$)
in the slave-particle representation of operators for atoms.
The usefulness of the slave-particle representation 
in various aspects has been
pointed out  for the original fermionic 
$t$-$J$ model\cite{CP1}. We expect that similar advantage
of the slave-particle picture holds also
in the  bosonic $t$-$J$ model\cite{sachdev}.

In the slave-particle representation,
the bosonic operator $B_{x\sigma}$ for atom at the site $x$ and spin
$\sigma = 1,2$ is viewed as a composite of a spinon $z_{x\sigma}$
and a holon $\phi_{x}$\cite{aoki}, 
\be
B_{x\sigma} = \phi^\dag_{x} z_{x\sigma},
\ee 
where $z_{x\sigma}$ is the CP$^1$ spin field (Schwinger boson)
and $\phi_x$ is the HCB.

In Ref.\cite{aoki}, we replaced  these HCB operators of holons 
$\phi_x$ by the Higgs field  
with a definite amplitude, 
\be
\phi_x \rightarrow \sqrt{\mathstrut \delta} \exp(i\varphi_x),
\ee
where $\delta$ is the average density of holes (holons) $\delta = 
\la \phi^\dag_x\phi_x\ra$ and $\varphi_x$ is the phase degrees of freedom.
This replacement is an approximation to ignore the fluctuation of
amplitude of holons assuming their {\it homogeneous distribution}.
By the MC simulations we have obtained a
phase diagram and various correlation functions of 
the 3D model at finite temperatures ($T > 0$)
and the 2D model at zero temperature ($T=0$).
In both cases, we found the coexistence region of AF order and SF.

It is then interesting to relax the above assumption of
homogeneous distribution of holes and consider the possibility
of PS.
With an isotropic AF coupling,
it is reported\cite{uniform} that the ground state of the
2D bosonic t-J model is spatially homogeneous
without PS for $J/ t \leq1.5$. 
In this paper, we shall study the 3D bosonic $t$-$J$ model
at finite $T$
in the slave-particle representation
{\it without} the above mentioned approximation.
That is, we treat the holon variables $\phi_x$ as genuine HCB
instead of the compact Higgs field $\exp(i\varphi_x)$.
We express  $\phi_x$ as another CP$^1$ (pseudo) spin field $w_{x\eta}$
$(\eta=1,2)$
via pseudo-spin SU(2) operator as 
\be
\phi_x = w^\dag_{x2} w_{x1}.
\label{phiw}
\ee
We study this double-CP$^1$ system of $z_{x\sigma}$ and $w_{x\eta}$
by  MC simulations. We examine its phase structure, various correlation 
functions, and possible PS, and so on.  

The present paper is organized as follows.
In Sec.2, we introduce the
double-CP$^1$ representation of the bosonic $t$-$J$ model 
in the 3D lattice at finite $T$'s.
We define two versions of the model, Model I directly derived from the 
$t$-$J$ model,
and Model II, a simplified version of Model I. They
have different weights in spin stiffness. By comparing the results
of these two models, one may obtain further understanding of the 
interplay of holons and spinons.   
In Sec.3, we exhibit the results of the numerical study 
 and the phase diagram of the simplified mode, Model II first.
We calculated the specific heat, the spin and atomic
correlation functions.
From these results,
we conclude that there exists a coexisting phase of AF long-range 
order and SF in a region of low-$T$ and intermediate hole concentrations.
In Sec.4, we consider Model I
and present the MC results. We compare the phase structures
of the two models. 
We also study the PS of both models.
Section 5 is devoted for conclusion.
In Appendix A, we derive the expression of the HCB
operator $\phi_x$ in terms of another CP$^1$ operator $w_{x\eta}$.

\section{Model I and Model II}
\setcounter{equation}{0} 

Let us start with the bosonic Hubbard model
of HCB's.
Its Hamiltonian is given by
\be
H_{\rm Hub}&=&-t\sum_{x,\mu,\sigma}\big(B^\dagger_{x+\mu,\sigma}
B_{x\sigma}+{\rm H.c.}\big)
+U\sum_x \hat{n}_{x1}\hat{n}_{x2},\nn
\hat{n}_{x\sigma}
&\equiv& B^\dag_{x\sigma}B_{x\sigma},
\label{Hhub}
\ee
where $B_{x\sigma}$ is the HCB operator\cite{B0,hcb}
to describe the bosonic atom at the site $x$ of the 3D cubic lattice
and the spin $\sigma [\ =1(\uparrow), 2(\downarrow)]$. 
$\mu (=1,2,3)$ is the 3D direction index and also denotes
the unit vector.
$B_{x\sigma}$ satisfies the following mixed commutation relations
for HCB's\cite{acr},
\be
\left[B_{x\sigma}, B^\dag_{x\sigma'}\right]_+ &=& \delta_{\sigma\sigma'},\ 
\left[B_{x\sigma}, B_{x\sigma'}\right]_+ = 0, \nn
\left[B_{x\sigma}, B^\dag_{y\sigma'}\right] &=& 
\left[B_{x\sigma}, B_{y\sigma'}\right] = 0\ {\rm for}\ x\neq y.
\label{B}
\ee  
We note that $H_{\rm Hub}$ has a global SU(2) symmetry under
\be
B_{x\sigma}\to B_{x\sigma}'=\sum_{\sigma'}
g_{\sigma\sigma'}B_{x\sigma'}, \ g \in\ {\rm SU(2)}.
\label{su2}
\ee
Actually, the first term is manifestly invariant.
To show that the second term is also invariant,
we introduce the atomic number operator,
\be
\hat{n}_x&\equiv&\hat{n}_{x1}+\hat{n}_{x2},
\ee
which is also invariant under (\ref{su2}).
Then the second term is expressed sorely by 
this  invariant quantity $\hat{n}_{x}$ as
\be
\hat{n}_{x1}\hat{n}_{x2}&=&\frac{1}{2}(\hat{n}_{x1}+\hat{n}_{x2})^2
-\frac{1}{2}(\hat{n}^2_{x1}+\hat{n}^2_{x2})\nn
&=&\frac{1}{2}(\hat{n}_{x}^2-\hat{n}_{x}),
\ee
where we used the relation $\hat{n}^2_{x\sigma}=\hat{n}_{x\sigma}$,
which holds because the  eigenvalue of $\hat{n}_{x\sigma}$ is 
0 or 1 due to HCB.

From this Hubbard model one may derive the bosonic $t$-$J$ model
as its effective model for strong correlations, i.e., 
for large $U$, and small hole concentrations.
The Hamiltonian $H$ of the 3D bosonic $t$-$J$ model is given by
\be
H&=&-t\sum_{x,\mu,\sigma}\big(\tilde{B}^\dagger_{x+\mu,\sigma}
\tilde{B}_{x\sigma}+{\rm H.c.}\big)\nn
&&+J\sum_{x,\mu}\left(\vec{\hat{S}}_{x+\mu}\cdot\vec{\hat{S}}_x
-\frac{1}{4}\hat{n}_{x+\mu}\hat{n}_x\right),\nn
\tilde{B}_{x\sigma} &\equiv&  (1-\hat{n}_{x\bar{\sigma}})\ B_{x\sigma},\ 
(\bar{1}\equiv 2, \bar{2}\equiv 1),\nn
\vec{\hat{S}}_x&\equiv&\frac{1}{2}B^\dagger_x \vec{\sigma}B_x,\ \ 
(\vec{\sigma}:{\rm Pauli\ matrices}).
\label{tJH}
\ee
The derivation of (\ref{tJH}) is achieved just by following the steps 
developed in the theory of high-$T_{\rm c}$ superconductivity
to derive the fermionic $t$-$J$ model from
the standard Hubbard model.
Actually, the second-order perturbation theory in terms of
small parameter $t/U$ produces (\ref{tJH})
with the relation $J = 4t^2/U$\cite{3sites}.  
The role of Pauli principle for fermions is played here 
by the hard-core nature of bosons.

Because of the hard-core nature of $B_{x\sigma}$, $(B_{x\sigma}^\dag)^2=0$,
the states of double occupancy of the same-spin bosons at each site 
are excluded.
Furthermore, as we consider the case of strong repulsion between 
two bosons with opposite spins at the same site, we
impose another constraint that excludes from the physical space
the double-occupancy state of two bosons with the opposite spins at each
site. This implies the states 
$B^\dagger_{x\uparrow} B^\dagger_{x\downarrow}|0\rangle$
are excluded from the physical space\cite{doubleoccupancy}. 
The tilde operator $\tilde{B}_{x\sigma}$ reflects this fact.
These two constraints are expressed for the normalized physical states
$|{\rm phys}\ra$ as
\be
\hspace{-0.5cm}
\la {\rm phys}|\hat{n}_x|{\rm phys}\ra=
\langle {\rm phys}|\sum_{\sigma}B^\dagger_{x\sigma}B_{x\sigma}
|{\rm phys}\rangle
\le 1.
\label{const0}
\ee

The Hamiltonian (\ref{tJH}) respects SU(2) symmetry (\ref{su2}).
To see it, it is helpful to rewrite $\tilde{B}_{x\sigma}
=(1-\hat{n}_{x\bar{\sigma}})B_{x\sigma}=(1-\hat{n}_x)B_{x\sigma}$
due to $B_{x\sigma}^2=0$. 

As mentioned, the  fermionic counterpart of the present bosonic model,
the (fermionic) $t$-$J$ model, is a 
canonical model
for strongly-correlated electron systems like the high-temperature
superconductors.
The 2D fermionic $t$-$J$ model has been studied intensively 
by means of various methods since the discovery of the high-$T$ 
superconductors, although its phase structure etc. 
are  still not clarified\cite{hightc}. The slave-particle approach like
the slave-boson or slave-fermion representations has provided
 us with an intuitive way of description of the $t-J$ model
 in the mean field theory, in the
 charge-spin separated state\cite{css}, and so on.
It is natural to expect that the slave-particle approach 
is useful also in the bosonic $t-J$ model.    

In the slave-particle representation, $B_{x\sigma}$
is expressed as
\be
B_{x\sigma} = \phi^\dagger_x a_{x\sigma}.
\ee 
$\phi_x$ represents the annihilation operator of the 
hard-core holon, and  satisfies the HCB algebra of single species,
\be
\left[\phi_{x}, \phi^\dag_{x}\right]_+ &=& 1,\ 
\left[\phi_{x}, \phi_{x}\right]_+ = 0, \nn
\left[\phi_{x}, \phi^\dag_{y}\right] &=& 
\left[\phi_{x}, \phi_{y}\right] = 0\ {\rm for}\ x\neq y.
\label{phi}
\ee  
$a_{x\sigma}$ represents annihilation operator of the bosonic 
spinon carrying $s=1/2$ spin. Their commutation relations are
\be
\left[a_{x\sigma}, a^\dag_{y\sigma'}\right] &=& \delta_{xy}
\delta_{\sigma\sigma'},\quad
\left[a_{x\sigma}, a_{y\sigma'}\right] = 0.
\label{a}
\ee
The physical-state condition (\ref{const0}) is replaced by the
following equation in terms of the slave-particle 
operators $a_{x\sigma}$
and $\phi_x$,
\begin{equation}
(\sum_\sigma 
a_{x\sigma}^\dagger a_{x\sigma} +\phi_x^\dag \phi_x)|{\rm phys}\rangle
 = |{\rm phys}\rangle.
\label{const1}
\end{equation}
Meaning of Eq.(\ref{const1}) is obvious.
There are three physical states for each $x$;
one-holon state (holon number $\phi_{x}^\dag \phi_x =1$), 
corresponding to the state with no atoms, 
and the two one-spinon states (one with spinon number $a_{x1}^\dag 
a_{x1} =1$ and the other with $a_{x2}^\dag a_{x2} =1$) 
corresponding to the two one-atom states.
The correspondence among these physical states are 
given as  follows:
\be
B_{x\sigma}|0\ra &=&0,\nn
 \phi_x|{\rm vac}\ra&=&a_{x\sigma}|{\rm vac}\ra=0,\nn
|0\ra &=& \phi_x^\dag|{\rm vac}\ra,\ \ {\rm No-atom\ state}\nn     
B_{x\sigma}^\dag|0\ra &=& a_{x\sigma}^\dag|{\rm vac}\ra,\ {\rm Two\ 
one-atom\ states}.
\label{physicalstates}
\ee

In terms of the slave-particle operators, 
the Hamiltonian (\ref{tJH}) becomes 
\begin{eqnarray}
H&=&-t\sum_{x,\pm\mu,\sigma}\phi_x^\dagger a_{x\pm \mu,\sigma}^\dagger 
a_{x\sigma}\phi_{x\pm \mu}
\nonumber \\
&&+{J \over 4}\sum_{x,\mu}\left[(a^\dagger\vec{\sigma}a)_{x+\mu}\cdot
(a^\dagger\vec{\sigma}a)_x-(a^\dagger a)_{x+\mu}(a^\dagger a)_{x}\right],
\nn
\label{tJH2}
\end{eqnarray}
where we write $(a^\dagger a)_x \equiv 
\sum_{\sigma}a^\dagger_{x\sigma} a_{x\sigma},\
(a^\dagger \vec{\sigma} a)_x \equiv \sum_{\sigma,\sigma'}
a^\dagger_{x\sigma}\vec{\sigma}_{\sigma\sigma'} a_{x\sigma'}$.

The constraint (\ref{const1}) is solved as
\be 
a_{x\sigma} = (1-\phi^\dag_x\phi_x)z_{x\sigma},
\label{az}
\ee
where $z_{x\sigma}$ is the CP$^1$ spin operator
that satisfies the usual commutation relations for bosons
and the following CP$^1$ constraint,
\be
\sum_{\sigma} z_{x\sigma}^\dag z_{x\sigma}=1.
\ee
To derive Eq.(\ref{az}) we have used the identity
\be
(1-\phi^\dag_x \phi_x)^2 &=& (1-\phi^\dag_x \phi_x),
\ee
because the eigenvalues of $\phi^\dag_x \phi_x$ are 0 and 1.
  
In Appendix A, we show that
the HCB operator $\phi_x$ can be exactly expressed 
in terms of another CP$^1$  operator $w_{x\eta}\ (\eta=1,2)$,
which satisfy the usual bosonic commutation relations and the constraint,
\be
\sum_{\eta=1}^2 w_{x\eta}^\dag w_{x\eta}=1.
\ee
The expression of $\phi_x$ is then given by 
\be
\phi_x = w^\dag_{x2} w_{x1}.
\label{phiw2}
\ee 
Then we employ the path-integral expression
of the partition function 
\be 
Z = {\rm Tr}\exp(-\beta H),
\label{Z}
\ee
at $T$ with $\beta \equiv 1/(k_B T)$.
For this purpose we introduce
two sets of CP$^1$ variables, $z_{x\sigma}(\tau), w_{x\eta}(\tau)$
at  each site $x$ and imaginary time 
$\tau \in [0,\beta\equiv (k_{\rm B} T)^{-1}]$.
(Hereafter we often set the Boltzmann constant $k_B$ to unity.)
They are complex numbers satisfying
\be
\sum_{\sigma=1}^2 \bar{z}_{x\sigma}(\tau) z_{x\sigma}(\tau) &=& 1,\nn
\sum_{\eta=1}^2 \bar{w}_{x\eta}(\tau) w_{x\eta}(\tau) &=& 1.
\label{z}
\ee
Then the partition function  is given by
means of path integrals over  $z_{x\sigma}(\tau)$ and
$w_{x\eta}(\tau)$.

To proceed further, we make one simplification
by considering finite-$T$ region, such that
the $\tau$-dependence of the variables in the path integral 
can be ignored keeping only the zero modes, i.e., $z_{x\sigma}(\tau)
\rightarrow z_{x\sigma}$, etc.\cite{aoki}.
To study the  phase structure at finite $T$ is very important because
it summarizes the essential properties of the system.
Besides it, finite-$T$ phase diagram gives a
very useful insight into the phase structure at $T=0$, i.e.,
if some ordered states are found at finite $T$, we can naturally expect that
they persist down to $T=0$.

Let us present the path-integral expression $Z_{\rm I}$
of the partition function (\ref{Z}) of the 3D model at finite $T$'s.
\be
Z_{\rm I}&=&\int \prod_{x}\big[dz_xdw_x
\prod_\mu dU_{x\mu}\big]\exp (A_{\rm I}-\mu_c\sum_x \bar{\phi}_x\phi_x),
\nn
\ee
where the suffix I has been attached because we call this model Model I
(We shall introduce Model II later).
The action $A_{\rm I}$ on the 3D lattice is given by
\begin{eqnarray*}
A_{\rm I}&=&A_{\rm s}+A_{\rm h},\nn
A_{\rm s}&=&\frac{c_1}{2}\sum_{x,\mu,\sigma}P_xP_{x+\mu}
\Big(\bar{\tilde{z}}_{x+\mu,\sigma}U_{x\mu}z_{x\sigma} + \mbox{c.c.}\Big), 
\nn
A_{\rm h}&=&{c_3 \over 2}\left[
\sum_{x,\mu,\sigma}\bar{z}_{x+\mu,\sigma}z_{x\sigma} 
\phi_{x+\mu}\bar{\phi}_x +{\rm c.c.}\right],
\end{eqnarray*}
\be
P_x &\equiv& 1-\bar{\phi}_x\phi_x,\nn
\tilde{z}_{x1}&\equiv&\bar{z}_{x2},\ \tilde{z}_{x2}\equiv-\bar{z}_{x1}\
(\tilde{z}_{x} \equiv i\sigma_2 \bar{z}^{\rm t}_x),\nn
\phi_x &\equiv&\bar{w}_{x2} w_{x1},\ U_{x\mu} \equiv \exp(i\theta_{x\mu}),
\label{action}
\ee
and the integration measure is
\be
\int dz_x &=&\int_{-\infty}^{\infty} 
dz_{x1}\int_{-\infty}^{\infty} 
dz_{x2}\ \delta(\sum_\sigma\bar{z}_{x\sigma}z_{x\sigma}-1) \ {\rm etc.},\nn
\int dU_{x\mu}&=& \int_0^{2\pi}\frac{d\theta_{x\mu}}{2\pi}.
\ee
Here we have used the same letter $\phi_x$ 
for the complex variable
and the operator in Eq.(\ref{phi}) because no confusions arise. 
According to Ref.\cite{aoki},
we have introduced the  U(1) gauge
field $U_{x\mu} \equiv \exp(i\theta_{x\mu})$ on the link $(x,x+\mu)$
as an auxiliary field to make the action in a simpler form
and the U(1) gauge invariance manifest.

The  term $-\mu_c\sum_x \bar{\phi}_x\phi_x$ with
the (minus of) chemical potential $\mu_c$ has been introduced
to control the hole density $\rho$,
\be
\rho(c_1,c_3,\mu_c)\equiv \frac{1}{N}\sum_{x}\la \bar{\phi}_x \phi_x \ra,
\ee
where $N=\sum_x 1$ is the total number of the sites.
As indicated, $\rho(c_1,c_3,\mu_c)$ is a function of $c_1, c_3$ and $\mu_c$.
We are interested in the case 
that the hole density $\rho$ takes a constant value $\delta$
as $c_1,c_3,\mu_c$ are varied,
because each material has a constant hole density, $\delta$. 
Then, to  obtain a physical quantity for given $c_1, c_3$ and $\delta$,
we determine $\mu_c$ so that the relation
\be
\rho(c_1,c_3,\mu_c) = \delta,
\ee  
holds.  It implies to determine $\mu_c(c_1, c_3,\delta)$
as a function of $c_1, c_3,\delta$.

Because the equation
\be
\hat{n}_x = 1-\phi^\dag_x\phi_x,
\label{nx}
\ee
holds for the physical states,
there is a relation among the atomic density $n$
and the hole density $\rho$ as
\be
n \equiv \frac{1}{N}\la \sum_x \hat{n}_x \ra = 1-\rho,
\ee
as expected.

The action $A_{\rm I}$ of (\ref{action}) 
has the global SU(2) spin symmetry of (\ref{su2}), 
\be
z_{x}&\rightarrow& g z_{x},\ \ g \in {\rm SU(2)}\nn
\bar{z}_{x+\mu} z_x &\rightarrow&
\bar{z}_{x+\mu}\bar{g}g z_x = \bar{z}_{x+\mu} z_x,\nn
 \bar{\tilde{z}}_{x+\mu} z_x
&\rightarrow& {\rm det}g\cdot \bar{\tilde{z}}_{x+\mu} z_x
= \bar{\tilde{z}}_{x+\mu} z_x.
\ee
We note that the present model may describe not only $s=1/2$ case
but also the case with an {\it arbitrary} $s$.
In this case, Eq.(\ref{z}) becomes as\cite{2s}
\be
\sum_{\sigma=1}^2 \bar{z}_{x\sigma}z_{x\sigma}=2s,
\label{2s}
\ee
but the above normalization factor $2s$ can be easily absorbed
into the parameters $c_1$ and $c_3$ in the action.

The action $A_{\rm I}$ is also invariant 
under a local ($x$-dependent) U(1) gauge transformation,
\be
z_{x\sigma} &\rightarrow& e^{i\lambda_x}z_{x\sigma},\ 
\phi_x \rightarrow e^{i\lambda_x} \phi_x,\nn
U_{x\mu}&\rightarrow& e^{-i\lambda_{x+\mu}}U_{x\mu}e^{-i\lambda_x},\nn
w_{x1}&\rightarrow& e^{i\frac{\lambda_x}{2}} w_{x1},\ 
w_{x2}\rightarrow e^{-i\frac{\lambda_x}{2}} w_{x2}, 
\ee
where $\lambda_x$ is an arbitrary function.
The complex variable
$B_{x\sigma}$\cite{B} for the gauge-invariant {\it bosonic} atoms
is expressed in terms of $z_\sigma$ and $w_\eta$ as 
\be 
B_{x\sigma} =
\bar{\phi}_x z_{x\sigma} = \bar{w}_{x1} w_{x2} z_{x\sigma}.
\label{b}
\ee
The parameters $c_1$ and $c_3$ are related with those in the
original $t$-$J$ model as\cite{aoki} 
\be
c_1&\sim& 
\left\{
\begin{array}{ll}
J\beta& {\rm for}\ c_1 >> 1,\\
(2J\beta)^{1/2}& {\rm for}\ c_1 << 1,
\end{array}
\right.
\nn 
c_3&\sim& t\beta.
\label{c1c3}
\ee
Note that $c_3$ here has no extra factor $\delta$
compared with $c_3$ defined in Ref.\cite{aoki}.

This model (\ref{action}), which we call Model I, 
is examined in Sect.4 as announced.
Before that, we study a simplified model, Model II, which
is defined 
by setting $P_x=1$ in (\ref{action}).
The partition function $Z_{\rm II}$ of Model II is then given by
\be
Z_{\rm II}&=&\int \prod_{x}\big[dz_xdw_x
\prod_\mu dU_{x\mu}\big]\exp (A_{\rm II}-\mu_c\sum_x \bar{\phi}_x\phi_x),\nn
A_{\rm II}&=&A'_{\rm s}+A_{\rm h},\nn
A'_{\rm s}&=&\frac{c_1}{2}\sum_{x,\mu,\sigma}
\Big(\bar{\tilde{z}}_{x+\mu,\sigma}U_{x\mu}z_{x\sigma} + \mbox{c.c.}\Big).
\label{action1}
\ee
The main reason to study Model II is 
to clarify the effect of the projection operator  to the
hole-free states, $P_x$,
by comparing the results of the two models.
We expect that the AF ordered state appears with stronger signals
in Model II than in 
Model I because the assignment $P_x=1$ in Model II
lets the AF coupling between the nearest-neighbor (NN) 
spin pairs at $(x,x+\mu)$  survive 
even if a hole occupies the sites $x$ and/or $x+\mu$. 
Here we note that, in the path-integral formulation, the variable
$z_{x}(\tau)$ is defined for all $(x,\tau)$ even if the site $x$ is
totally occupied by a hole\cite{FN1}. 
Due to the short-range AF configuration, such $z_x$ reflects
a nearby spin orientation.
In other words, Model II effectively describes a doped AF Heisenberg
model with, e.g., next-NN exchange couplings that enhance AF 
long-range order.

\section{Results of MC Simulations of Model II}
\setcounter{equation}{0} 

In this section, we 
present the results of MC simulations for Model II of (\ref{action1}).
For MC simulations, we consider a 3D cubic lattice 
of the size $N\equiv L^3$ ($L$ up to 36)
and 
imposed the periodic boundary condition.
We used the standard Metropolis algorithm with local updates.
Average number of sweeps was $12\times10^{4}$, and average 
acceptance ratio was about 40\%$\sim$70\%.

\subsection{Phase structure}

Let us first discuss the phase structure.
To this end, we measured the internal energy $U$,
the specific heat $C$ and the hole density $\rho$ defined as
\be
U&=&-\frac{1}{N}\langle A\rangle,\nn
C&=&\frac{1}{N}\left(\langle A^2\rangle 
-\langle A\rangle^2\right),\nn
\rho&=&\frac{1}{N}\langle\sum_{x}\overline{\phi}_{x}
\phi_{x}\rangle.
\label{UC}
\ee

\begin{figure}[t]
\begin{center}
\hspace{-0.5cm}
\includegraphics[width=9.0cm]{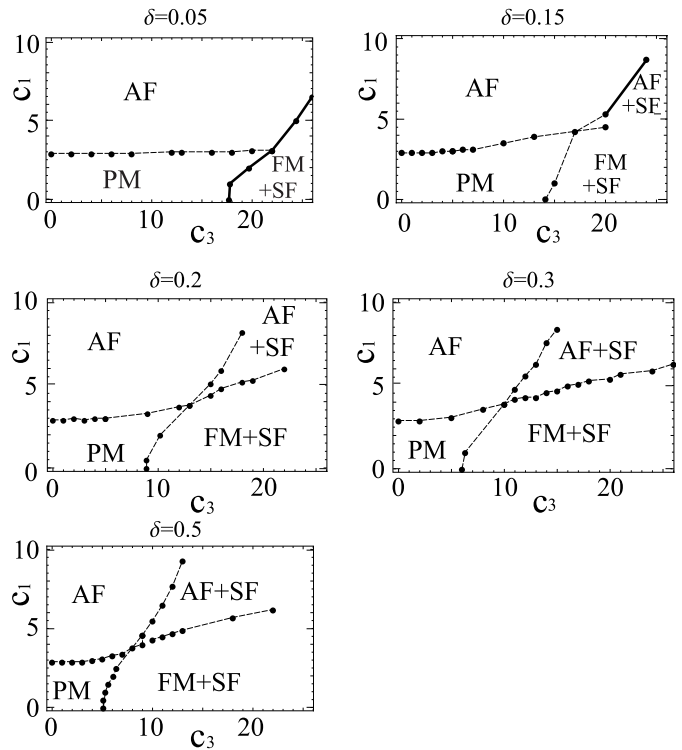}
\end{center}
\caption{
Phase structure of Model II in the $c_3$-$c_1$ plane for various $\delta$.
The meaning of abbreviations  for each phase is listed in
Table 1. The phase transition into the FM+SF phase for $\delta=0.05$
and the phase transition between the AF and AF+SF phases for $\delta=0.15$
(solid curves) are of first order. All the other transitions 
(dashed curves) are of second order.}
\label{fig:phase}
\end{figure}

In Fig.\ref{fig:phase} we show 
the obtained phase diagram in the $c_3$-$c_1$ plane for various values
of $\rho = \delta$.
The phase transition lines in Fig.\ref{fig:phase} were determined by 
the measurement of $U$ and $C$ in (\ref{UC}).
We found that there are  four phases listed in Table 1.
The nature of each phase was confirmed by
measuring various correlation functions, which we shall discuss
in the following subsection in detail.

Let us present a qualitative understanding 
of the phase diagram Fig.\ref{fig:phase}. 
For this purpose, it is convenient to
 introduce
an O(3) spin vector $\vec{\ell}_x$ made of spinon $z_x$,
\be
\vec{\ell}_x \equiv \bar{z}_x\vec{\sigma}z_x,\ \vec{\ell}_x\cdot\vec{\ell}_x=1,
\label{ell}
\ee
where summations over spin indices are understood.
The short-range (SR) spin correlation
$\vec{\ell}_{x+\mu}\cdot\vec{\ell}_x$ is expressed by the SR
CP$^1$ amplitude (such as $\bar{z}_{x+\mu}z_{x}$) as
\be
\vec{\ell}_{x+\mu}\cdot\vec{\ell}_x = 2|\bar{z}_{x+\mu}z_{x}|^2 -1
= -2|\bar{\tilde{z}}_{x+\mu}z_{x}|^2 +1,
\ee

\vspace{0.5cm}

\begin{tabular}{ll}
\hline
AF:&  Antiferromagnetic phase \\
PM:&  Paramagnetic phase \\
AF+SF:& Phase of antiferromagnetism and SF\\
FM+SF:& Phase of ferromagnetism and SF\\
\hline
\end{tabular}
\begin{center}
Table 1. Various phases in Fig.\ref{fig:phase} 
\end{center}

\noindent
where $z_x$ satisfies
\be
|\bar{z}_{x+\mu}z_{x}|^2+|\bar{\tilde{z}}_{x+\mu}z_{x}|^2 =1.
\ee
So the SRAF configuration $\vec{\ell}_{x+\mu}\cdot\vec{\ell}_x \simeq -1$
corresponds to $|\bar{\tilde{z}}_{x+\mu}z_{x}| \simeq 1$,
and the SRFM configuration
$\vec{\ell}_{x+\mu}\cdot\vec{\ell}_x \simeq 1$
corresponds to $|\bar{z}_{x+\mu}z_{x}| \simeq 1$.
The $c_1$ term $A'_{\rm s}$ of (\ref{action1}) controls the spin fluctuations 
and the  AF order ($|\bar{\tilde{z}}_{x+\mu}z_x| \simeq 1$) 
is generated for sufficiently large values of $c_1$.
The $c_3$ term $A_{\rm h}$ of (\ref{action}), (\ref{action1}) controls the rate 
of hopping of holons, $\phi^\dag_{x+\mu}\phi_x$,
accompanied with a SRFM spinon amplitude $z_{x+\mu}\bar{z}_x$.
The SF of holons is generated 
for sufficiently large values of $c_3$ if there exists a sufficient
amount of FM spin amplitude $\langle z_{x+\mu}\bar{z}_x\rangle$.
Therefore, for sufficiently large values of $c_3$, we expect
a coherent SF of holons accompanied with a (SR) FM order.
As the results in Fig.\ref{fig:phase} show, there exists the coexisting
phase of AF and SF for intermediate and large hole density.
More details of the spin configuration in this phase will be
given in the following sections.

\begin{figure}[t]
\begin{center}
\includegraphics[width=8.5cm]{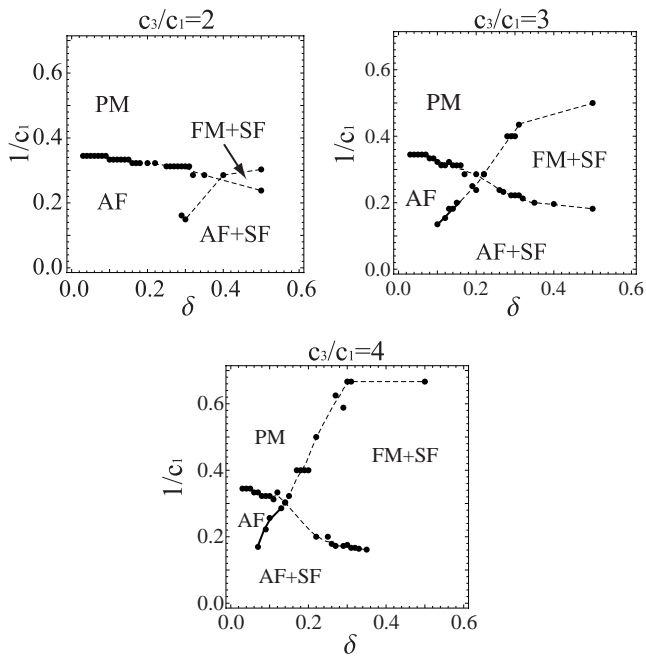}
\end{center}
\vspace{-0.5cm}
\caption{
Phase structure of Model II in the $\delta$-$1/c_1$ plane 
($1/c_1 \sim k_B T/J$) for $c_3/c_1 (\sim t/J) = 2.0,3.0,4.0$.
The abbreviations  for each phase are listed in Table 1. 
The phase transition between the AF and AF+SF phases for 
$c_3/c_1=3.0,4.0$ drawn in solid curves 
are of first order. All the other transitions drawn in
dashed curves are of second order.}
\label{fig:phasev2}
\end{figure}

In Fig.\ref{fig:phasev2} we show the phase diagram
drawn in the $\delta-1/c_1$ plane for several values of $c_3/c_1$.
The value for vertical axis, $1/c_1$,  and 
the ratio $c_3/c_1$ are expressed 
for $c_1 >> 1$  from (\ref{c1c3}) as 
\be
\frac{1}{c_1} &\sim& \frac{1}{J\beta} 
 \sim \frac{k_B}{J} T,\ \ 
\frac{c_3}{c_1} \sim \frac{t}{J}.
\ee
So Fig.\ref{fig:phasev2} may be interpreted as the phase diagram
in the $\delta-T$ plane\cite{deltaT}.
The N\'eel temperature $T_{\rm N}(\delta)$ 
(the critical temperature of the AF transition) decreases
as holons are doped (as $\delta$ increases).
The rate of reduction  $-dT_{\rm N}(\delta)/d\delta$ increases 
as $t/J$ increases
because the movement of holons becomes significant to destroy AF order.
The critical temperature $T_{\rm SF}(\delta)$ for SF of holons
increases as $\delta$ increases as expected. 
The rate of increase $dT_{\rm SF}(\delta)/d\delta$ also rises
as $t/J$ increases, because the movement of holons favors
their homogeneous distribution and so their SF.

\begin{figure}[t]
\begin{center}
\hspace{-0.5cm}
\includegraphics[width=8.5cm]{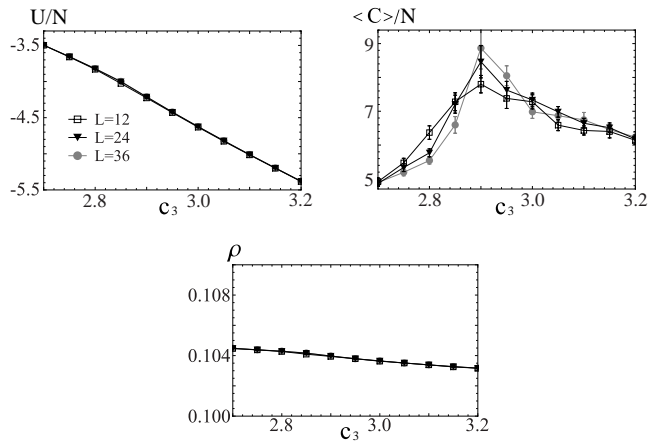}
\end{center}
\caption{
$E$, $C$ and $\rho$ along $c_3=4.0$ and $\delta \simeq 0.10$.
A transition between the PM and AF phases exists
at $c_1 \simeq 2.9$. Corresponding value of chemical
potential $\mu_c=10$. 
}
\label{fig:af2pm}
\end{figure}

Let us examine some details of the phase transitions in Fig.\ref{fig:phase}.
First, we focus on the transition from the PM phase to the AF phase
that takes place as the value of $c_1$ is increased. 
In Fig.\ref{fig:af2pm} we present $U$, $C$ and $\rho$ along $c_3=4.0,$
$\delta \simeq 0.10$  to locate the transition point
between the PM and AF phases.  
It seems that the phase transition is of second order 
because there are no hysteresis in $U$ and the peak of $C$
develops systematically as $L$ increases.
We obtained similar behavior of $U$ and $C$ in the PM-AF
phase transition for cases of other values of $\delta$ shown
in Fig.\ref{fig:phase}.

\begin{figure}[b]
\begin{center}
\hspace{-0.5cm}
\includegraphics[width=8cm]{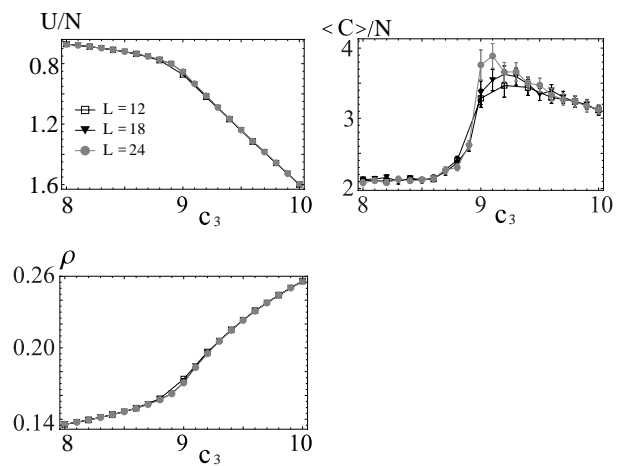}
\end{center}
\caption{
$U$, $C$ along $c_1=1.5$ and $\mu_c=10.0$.
A second-order transition between the PM and FM+SF phases exists
at $c_3 \simeq 9.0$.
}
\label{fig:pm2fmbc}
\end{figure}

Secondly, we examine the transition between the PM phase 
and the FM+SF phase. 
In  Fig.\ref{fig:pm2fmbc}
we present $U,C$ along $c_1=1.5$ and $\mu_c=10.0$, which indicate
a second-order transition at $c_3 \simeq 9.0$ between
the PM and FM+SF phases. 
$\rho$ increases as $c_3$ increases crossing the transition point.

\begin{figure}[t]
\begin{center}
\hspace{-0.5cm}
\includegraphics[width=8cm]{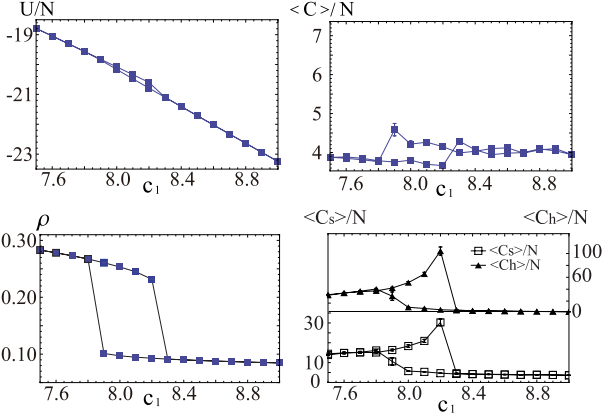}
\end{center}
\vspace{-0.5cm}
\caption{
$U$, $C$, $\rho$  vs $c_1$ for $c_3=24.0$ and $\mu_c=16.0$ $(L=12)$.
A first-order transition between the AF and AF+SF phases exists
at $c_1 \simeq 8.2$.
}
\label{fig:af2afbc}
\end{figure}

Thirdly, we examine the transition between the AF and AF+SF phases
at $\delta \simeq 0.15$. In Fig.\ref{fig:af2afbc} we show
$U$, $C$, $\rho$ for $c_3=24.0$ and $\mu_c=16.0$.
There are hysteresis curves in these quantities, which 
indicate a first-order phase transition.
Similar first-order transitions are found also
 for $\delta=0.05$
and $\delta=0.15$ as indicated by solid curves in Fig.\ref{fig:phase}.
We determined the transition point from the hysteresis data 
by averaging the two values of $\rho$ for fixed $\mu_c, c_1, c_3$, 
$\rho_1$ on the upper hysteresis curve and $\rho_2$
on the lower hysteresis curve,  so that
\be
a_1 \rho_1 + a_2 \rho_2 = \delta,\quad a_1+a_2=1,
\label{hysteresis}
\ee
to obtain the given value of $\delta$. $U$ and $C$ are averaged
with these weights $a_1, a_2$.  

\begin{figure}[b]
\begin{center}
\hspace{-0.5cm}
\includegraphics[width=8cm]{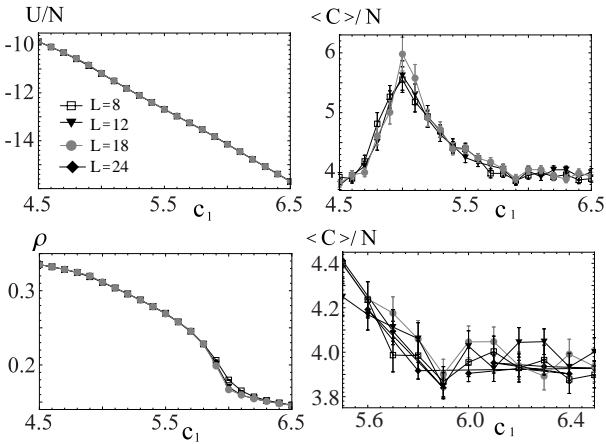}
\end{center}
\caption{
$U$, $C$, $\rho$  along $c_3=16.0$ and $\mu_c=10.0$.
A second-order transition between the FM+SF and AF+SF phases exists
at $c_1 \simeq 5.0$. There is a dip in $C$ at $c_1\simeq 5.9$,
which suggests another transition (See Fig.\ref{fig:chcs} and the text).
}
\label{fig:fmbc2afbc2af}
\end{figure}

\begin{figure}[t]
\begin{center}
\includegraphics[width=8cm]{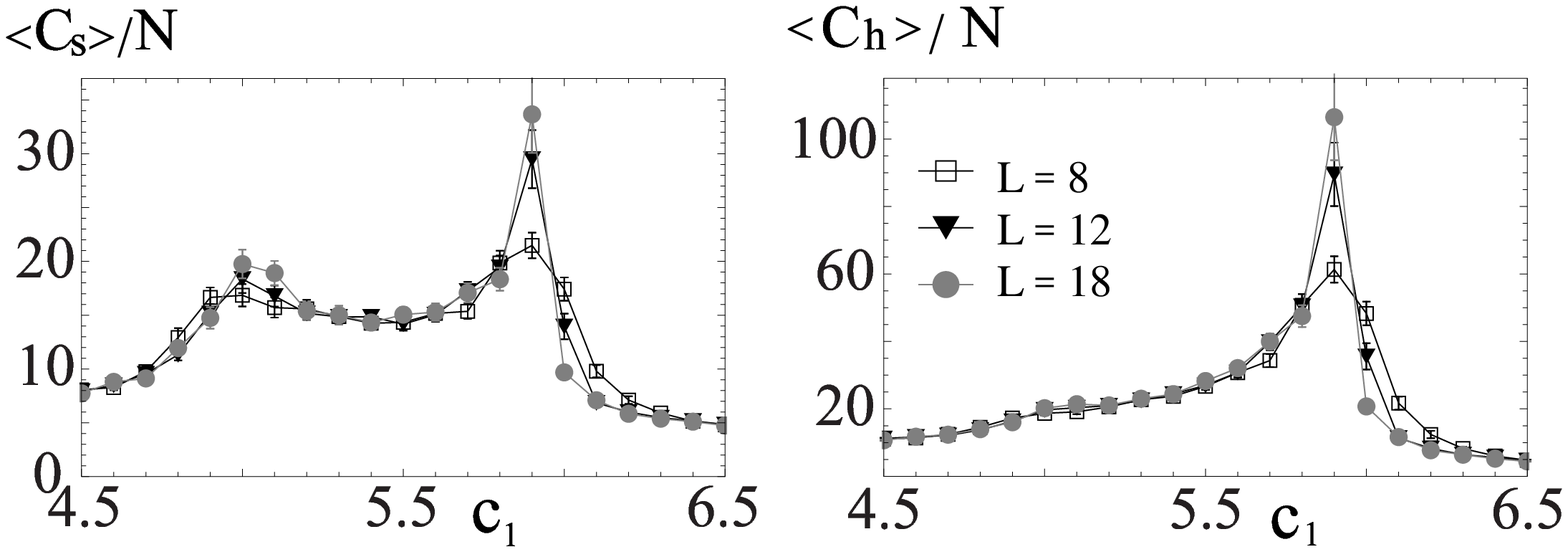}
\end{center}
\caption{
The specific heat for each term, $C_s$, $C_h$ along $c_3=16.0$
and $\mu_c=10.0$. They exhibit the corresponding peaks(or cusp)
for the transition between the FM+SF  and AF+SF phases
and the transition between the AF+SF and  AF phases.
}
\label{fig:chcs}
\end{figure}

Finally, we examine the successive transitions from
 the FM+SF phase to the AF+SF phase and then to the AF phase.
In Fig.\ref{fig:fmbc2afbc2af} we present
$U, C$ along $c_3=16.0$ and $\mu_c=10.0$.
We see that there is only one  peak in $C$ at $c_1\simeq 5.0$
that corresponds to the transition between
the FM+SF phase and the AF+SF phase,
the peak for the transition from the AF+SF phase to the AF phase
being missing.
(For identification of these phases, see later discussion.) 
However the hole density $\rho$ changes its behavior at 
$c_1\simeq 5.9$, while  $C$ exhibits only very small anomalous
behavior(dip) at that point.

To clarify the possible transition at $c_1\simeq 5.9$, we
measured the ``separate specific heats" for the $A'_{\rm s}$ term and the
$A_{\rm h}$ term separately. They are defined as
\be
C_{\rm s}&=&
\frac{1}{N}\left(\langle A_{\rm s}^2\rangle 
-\langle A_{\rm s}\rangle^2\right),\nn
C_{\rm h}&=&
\frac{1}{N}\left(\langle A_{\rm h}^2\rangle -\langle A_{\rm h}\rangle^2\right).
\ee

In Fig.\ref{fig:chcs} we present $C_{\rm s}$ and $C_{\rm h}$ separately.
Both $C_{\rm h}$ and $C_{\rm s}$ 
exhibit a very sharp peak at $c_1\simeq 5.9$.
This kind of ``cancellation" of  
specific heats of each term in the total specific heat $C$ 
has been observed sometimes.
This stems from the fact that $\langle A'_{\rm s}\rangle$ is an 
increasing function of $c_1$ whereas $\langle A_{\rm h}\rangle$ 
is a decreasing function of $c_1$.
At $c_1\simeq 5.9$, cancellation between these two terms occurs and 
$\langle A_{\rm II}\rangle$ becomes a smooth function of $c_3$.
The study of correlation function in the next subsection shows
a possible phase transition between AF+SF and AF phases at $c_1\simeq 5.9$. 

\subsection{Correlation functions}

In the previous subsection, we showed the phase diagrams 
for various hole density.
In order to understand physical properties of each phase,
let us study some correlation functions.
Because the spin operator   
$\vec{\hat{S}}_x$ of atom  
is given by Eq.(\ref{tJH}) as
\be
\vec{\hat{S}}_x = \frac{1}{2}B^\dag_x\vec{\sigma}B_x
= \frac{1}{2}(1-\phi^\dag_x\phi_x){z}^\dag_x\vec{\sigma}z_x,
\ee
we define the corresponding normalized classical atomic spin vector, 
\be
\vec{S}_x \equiv  
(1-\bar{\phi}_x\phi_x)\vec{\ell}_x,
\label{ell&S}
\ee
using the pure O(3) spin vector 
$\vec{\ell}_x =\bar{z}_x\vec{\sigma}z_x$ of Eq.(\ref{ell}).
We also introduce the  atomic density  $n_x$ corresponding to Eq.(\ref{nx}),
\be
n_{x}\equiv 1-\bar{\phi}_x\phi_x.
\ee
Then we measure the following correlation functions,
\be
G_{\rm \ell}(r)&=&\frac{1}{3N}\sum_{x,\mu}\left\langle
\vec{\ell}_{x+r\mu}\cdot\vec{\ell}_{x}\right\rangle,  \nn
G_{B}(r) &=&\frac{1}{6N(1-\delta)}\sum_{x,\mu,\sigma}\left\langle 
\bar{B}_{x+r\mu,\sigma} B_{x\sigma} \right\rangle,\nn
G_{n}(r) &=&\frac{1}{3N(1-\delta)^2}\sum_{x,\mu}\left\langle 
n_{x+r\mu} n_{x} \right\rangle,\nn
G_{S}(r) &=&\frac{1}{3N(1-\delta)^2}\sum_{x,\mu}\left\langle  
\vec{S}_{x+r\mu}\cdot \vec{S}_{x} \right\rangle. 
\label{corr}
\ee
From above, they are the spin correlation of spinons, 
atomic correlation, 
density correlation, and the spin correlation of atoms, respectively.
Their prefactors are chosen so that $G(r)$ is normalized as $G(0)=1$.

In Table 2, each phase is characterized by the (non)vanishing off-diagonal
long-range orders (LRO) $G(\infty)$ and/or 
staggered magnetizations  $\tilde{G}(\infty)$,
\be
G(\infty)&\equiv& \lim_{r \rightarrow \infty}G(r),\nn
\tilde{G}(\infty)&\equiv& \lim_{r \rightarrow \infty}(-)^r G(r),
\ee
measured by these correlation functions.

\begin{figure}[tb]
\begin{center}
\includegraphics[width=7cm]{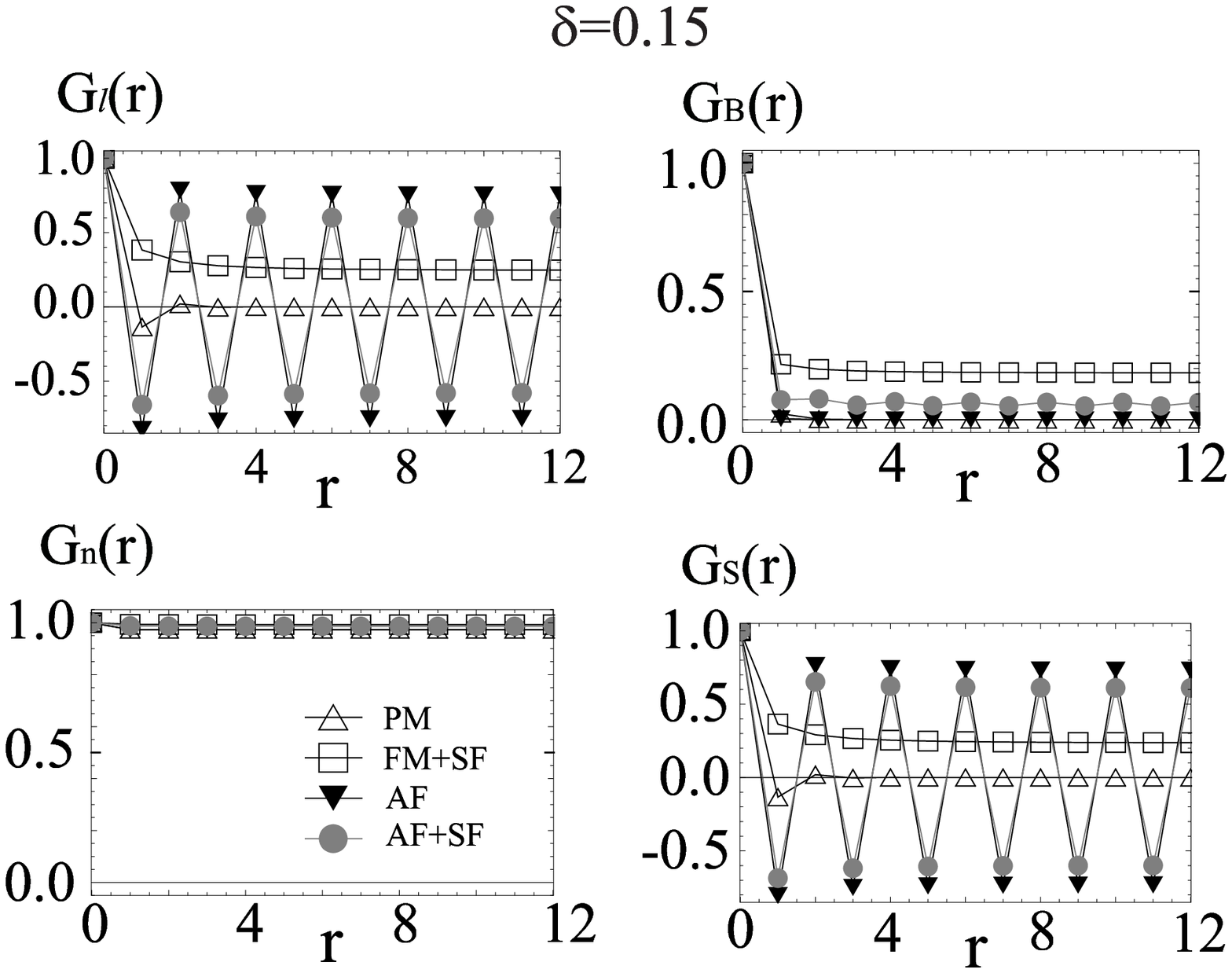}
\includegraphics[width=7cm]{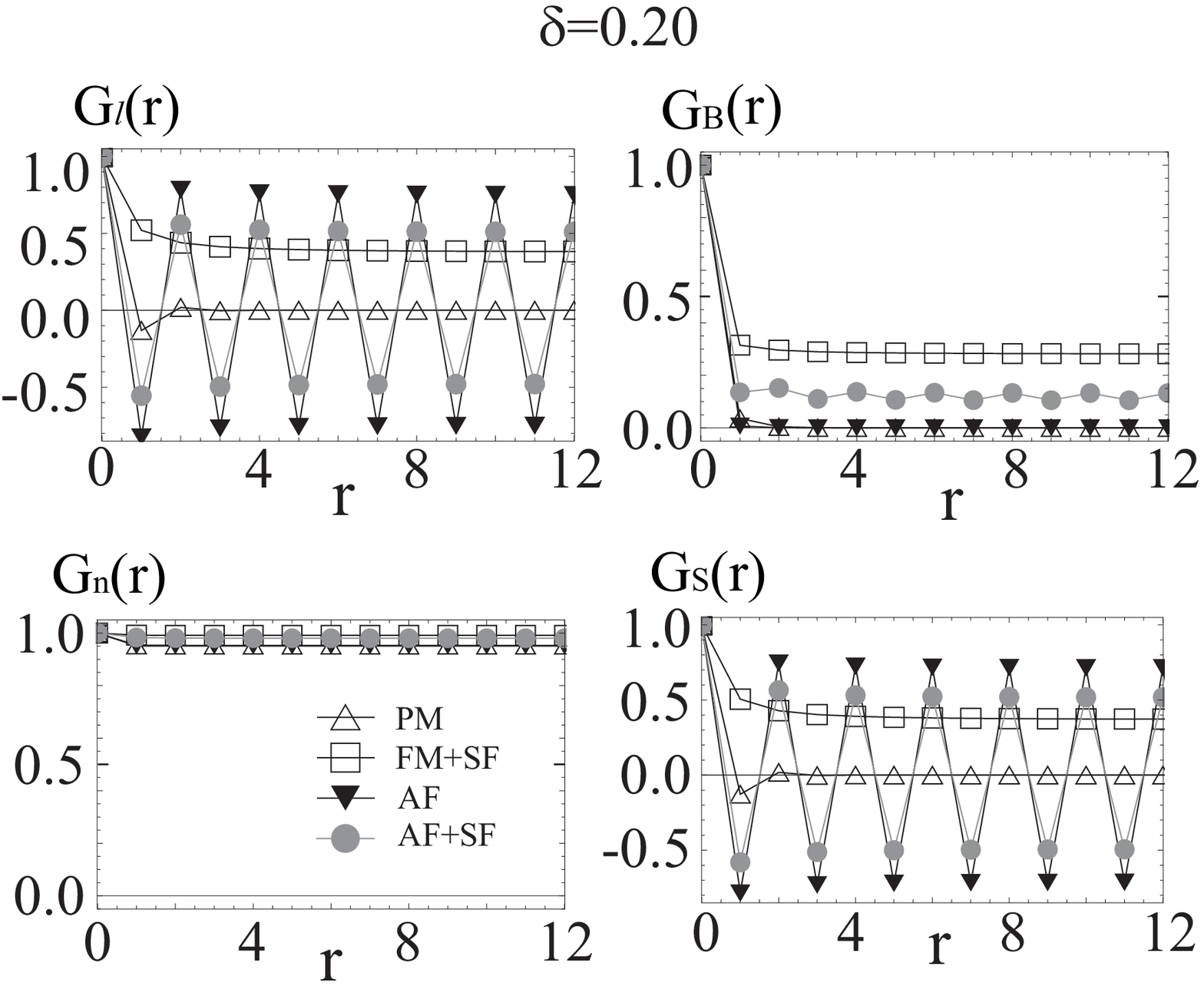}
\includegraphics[width=7cm]{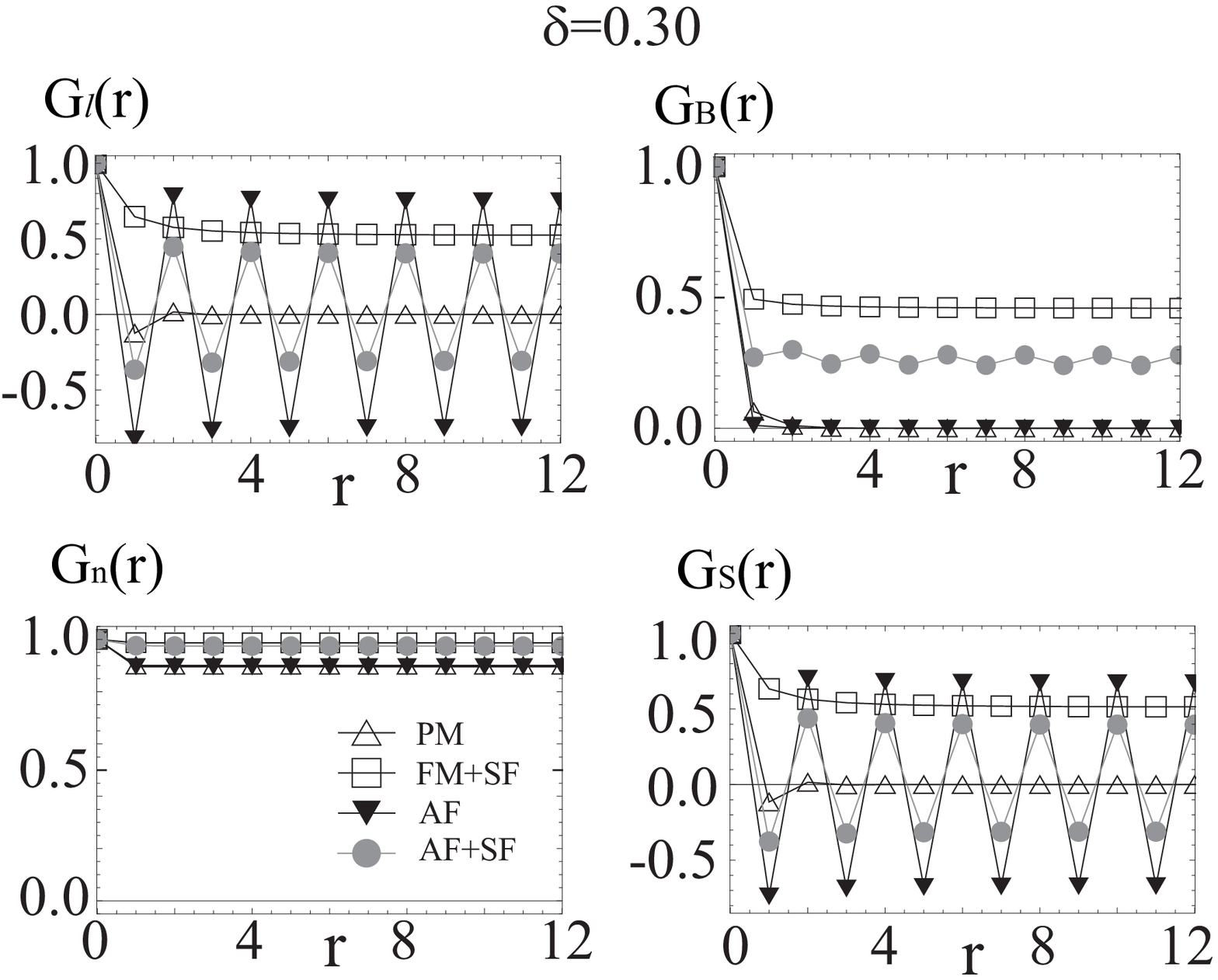}
\end{center}
\vspace{-0.5cm}
\caption{
Four correlation functions (\ref{corr})
at the four selected points in Table 3 
for $\delta=0.15, 0.20$ and $0.30$. They are used to
identify the phases of Fig.\ref{fig:phase}
according to Table 2.
}
\label{fig:cf}
\end{figure}

To show these correlation functions, 
we select a typical point 
in the $c_3$-$c_1$ plane for each phase as in Table 3.
In Fig.\ref{fig:cf} we present the four correlation functions
of (\ref{corr}) at the four selected points in Table 3
for three typical values of $\delta=0.15, 0.20$ and  $0.30$.
From this result, we identified each phase as in Fig.\ref{fig:phase}.
In the FM+SF phase and the AF+SF phase, the region
$0.08\alt\delta\alt 0.2$ requires a difficult fine tuning of $\mu_c$.
So, to obtain the correlation data for $\delta=0.15$ in these two phases,
we used the superposition (\ref{hysteresis}) by using the data for  
$\delta\simeq 0.08$ and $\delta\simeq0.2$.

The results in Fig.\ref{fig:cf} shows that the spin correlation 
$G_{\ell}(r)$ and the atomic-spin correlation $G_{S}(r)$
exhibit similar behavior in all phases.
They have staggered magne-

\vspace{0.3cm}
\begin{center}
\begin{tabular}{|c|c|c|c|}
\hline
     &$G_{\ell, S}(\infty)$&$G_{B}(\infty)$&$G_{n}(\infty)$
     \\\hline
PM   &$ =   0      $&$ =   0$&$\neq 0$ \\ \hline
FM+SF&$\neq 0\ $&$\neq 0$&$\neq 0$ \\ \hline
AF   &SM      &$ =   0$&$\neq 0$ \\ \hline
AF+SF&SM &$\neq 0$&$\neq 0$ \\ \hline
\end{tabular}
\end{center}
\nin
Table 2. (Non)Existence of LRO in each phase.
``SM" implies a nonvanishing staggered magnetization,
$\tilde{G}(\infty)$.\\

\begin{center}
\begin{tabular}{p{2.2cm}p{1.5cm}p{1.5cm}}
phase   &$c_3$&$c_1$\\
\hline
\hline
PM   &4.0&2.0\\
\hline
FM+SF&24.0&2.0\\
\hline
AF   &4.0&8.0\\
\hline
AF+SF&24.0&8.0\\
\hline
\end{tabular} \\
\vspace{0.3cm}
\end{center}
Table 3. Data points of correlation functions for each phase in the $c_3$-$c_1$ plane.


\nin
tizations  in the AF and AF+SF phases.
The atomic correlation $G_{B}(r)$ has nonvanishing
LRO in the FM+SF and AF+SF phases, which indicates the SF of atoms.
In particular, in the AF+SF phase, $G_{B}(r)$ of even-even (odd-odd)
sites is sightly different from that of even-odd sites,
reflecting the AF nature.
The atomic density correlation $G_n(r)$ shows smooth and homogeneous
distribution of atoms and holes at all the examined 
cases of $c_1, c_3, \delta$, indicating nonexistence of the
PS phenomenon in the present model.
This point will be discussed rather in detail in the following section.

\begin{figure}[t]
\begin{center}
\hspace{-0.2cm}
\includegraphics[width=7.7cm]{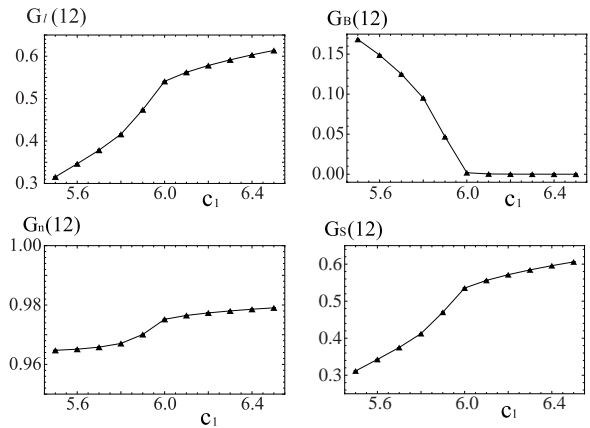}
\end{center}
\vspace{-0.5cm}
\caption{
Four correlation functions (\ref{corr}) at the long distance limit
$G(L/2)$ vs $c_1$ for $c_3=16.0$ and $\mu_c= 10.0$.
These values correspond to 
Figs.\ref{fig:fmbc2afbc2af},\ref{fig:chcs}. $G_B(L/2)$
shows a behavior of a second-order phase transition 
at $c_1\simeq 5.9\sim 6.0$.
}
\label{fig:cf2}
\end{figure}

\begin{figure}[b]
\begin{center}
\hspace{-0.5cm}
\includegraphics[width=7.7cm]{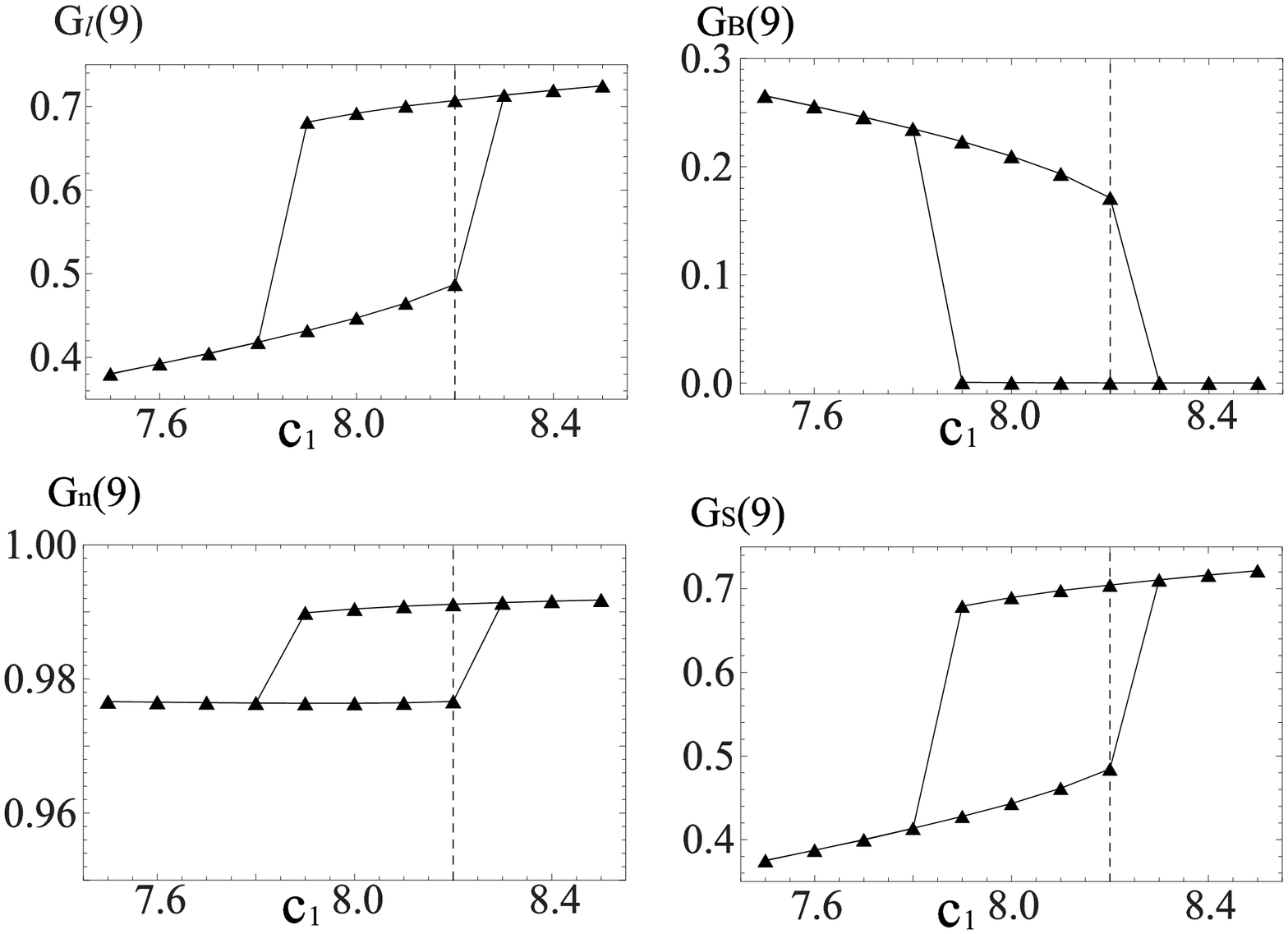}
\end{center}
\vspace{-0.5cm}
\caption{
Four correlation functions (\ref{corr}) at the long distance limit
$G(L/2)$ vs $c_1$ for $c_3=24.0$ and $\mu_c= 16.0$.
These values correspond to 
Fig.\ref{fig:af2afbc}. $G_B(L/2)$
shows a behavior of a first-order phase transition 
at $c_1\simeq 8.2$ (dashed line), which is determined from the specific heat
of Fig.\ref{fig:af2afbc} using (\ref{hysteresis}).
}
\label{fig:cf3}
\end{figure}

To study the possible phase transition between AF+SF and AF phases,
we measured the correlation functions across the relevant region of $c_1$.
In Fig.\ref{fig:cf2} we plot their values at long-distance limit,
$G(L/2)$ for $c_3=16.0$ and $\mu_c= 10.0$. 
Because $G_B(L/2)$  decreases continuously as $c_1$ increases
and vanishes at $c_1\simeq 5.9\sim 6.0$, we judge that
there is a second-order phase transition from 
the AF+SF phase to the AF phase. Other three values $G(L/2)$
remain finite as expected from Table 2.
Similarly, Fig.\ref{fig:cf3} shows  $G(L/2)$ for $c_3=16.0$
and $\mu_c = 16.0$.  Again $G_B(L/2)$ decreases  and vanishes
but discontinuously with hysteresis. This supports
our phase diagram, Fig.\ref{fig:phase}, 
with the first-order AF+SF $\leftrightarrow$ AFtransition
 at $c_1 \simeq 8.2$. 

\section{Results of MC Simulations of Model I}
\setcounter{equation}{0} 

In this section, we shall study  Model I of (\ref{action}).
Physical meaning of the prefactor 
$P_x=1-\bar{\phi}_x\phi_x$ 
in $A_s$ of (\ref{action}) is obvious, i.e., if the site $x$ is occupied by 
a holon, the AF coupling on the link $(x,x\pm\mu)$ is suppressed. 
This bond-breaking effect of holons in the 
AF background generates an effective attractive force 
favoring holon pairs sitting at NN sites. Actually,
a NN holon pair breaks eleven AF bonds while
two holons separated by more than NN sites break twelve AF bonds
\cite{2d3d}.
This effective force
can be an origin of the superconducting NN
holon-pair condensation\cite{CP1} and/or of a PS of
holon-rich regions and holon-free (holon-poor) regions.
In this section, we shall study how the phase diagram is
influenced by this AF bond-breaking effect by holons, in particular,
if there appears a phase-separated state.

\begin{figure}[b]
\begin{center}
\begin{tabular}{cc}
\hspace{-0.2cm}\includegraphics[width=9cm]{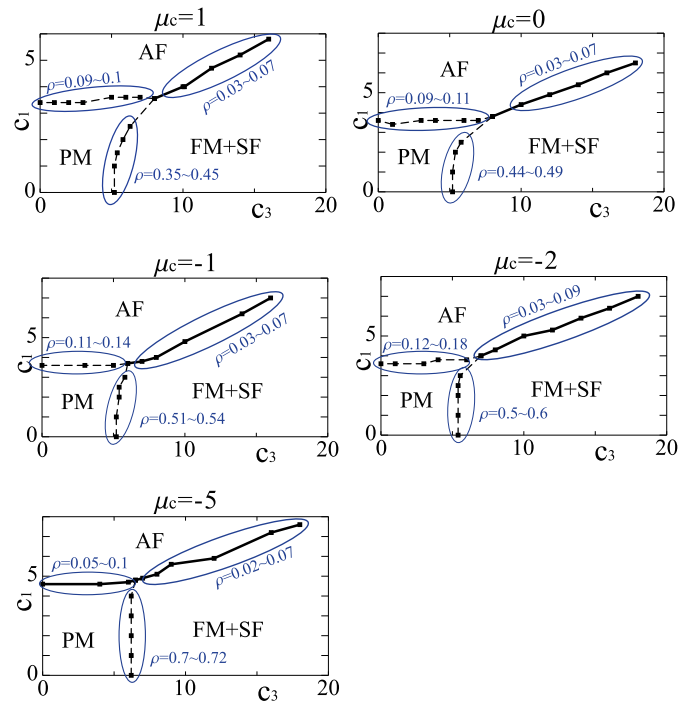}
\end{tabular}
 \end{center}
 \caption{Phase diagram of Model I in the $c_3-c_1$ plane for fixed values of
 the chemical potential,
 $\mu_c =+1.0,0.0,-1.0,-2.0,-5.0$. Hole density $\rho$
 is also shown for corresponding regions. The solid curves show
first-order transitions and the dashed curves show second-order transitions.
The AF+SF phase is missing.}
\label{PDM2} 
\end{figure}

\subsection{Phase structure}

As for Model II, we studied Model I by means of the MC simulations.
In Figs.\ref{PDM2}, we show the obtained phase diagram 
in the $c_3-c_1$ plane for fixed values of  $\mu_c$.
There exist the AF, PM and FM$+$SF phases as in
Model II with low-hole doping.
In the phase diagram, we also indicate the hole density $\delta$.
We found that hole density in the AF phase is rather small compared 
with Model II.
This is due to the factor $P_x$ in the $c_1$-term of (\ref{action}). 
As explained in the beginning of this section,
a hole sitting on the site $x$ makes $P_x=0$ and 
suppresses
the AF couplings around $x$. So
hole doping costs  larger energy in Model I than in 
Model II.
In other words, as the hole density is increased, the AF state
becomes more unstable because the AF couplings are weakened by
the prefactor $P_{x+\mu}P_x$.
By the numerical studies of Model I, we have concluded that 
the coexisting phase AF+SF of AF order and SF 
does not exists in Model I.

Order of each phase transition is indicated in Fig.\ref{PDM2}.
In particular, the transition between the AF and FM+SF is of
first order.
This result is physically expected as these two phases have
different LRO's in contrast.
We also found that the spin correlation $G_{\ell}$, atomic
correlation $G_B$, etc. exhibit a similar behavior to those in Model II.

\subsection{Phase separation and hole distribution}

The PS is a phenomenon of inhomogeneous distribution of holes
in which the lattice is separated into hole-rich regions 
and spin-rich (hole-free) regions.  
To study the possibility of PS, we introduced the 
following two quantities,
\be
\Delta&=&\frac{1}{N \rho^2}
\sum_x\left\langle \left(\bar{\phi}_{x}\phi_{x}-
\rho \right)^2\right\rangle,\nn
\Delta_{\rm c}&=&
\frac{1}{3N \rho^2}\sum_{x,\mu}\left\langle 
(\bar{\phi}_{x+\mu}\phi_{x+\mu}-\rho)(\bar{\phi}_{x}\phi_{x}-\rho)
 \right\rangle\nn
&=&\frac{1}{3N \rho^2}\sum_{x,\mu}\left\langle \bar{\phi}_{x+\mu}
\phi_{x+\mu}\bar{\phi}_{x}\phi_{x}-\rho^2 \right\rangle.
\label{psindex}
\ee
The quantity $\Delta$ measures the fluctuation of 
hole density $\bar{\phi}_x\phi_x$ around its average $\rho$. 
As explained in (\ref{physicalstates}), 
a physical state at a specific site $x$ is  
a superposition of the 
one-holon state $|0\rangle\ (B_{x\sigma}|0\ra = 0)$ and 
the no-holon states $B^\dag_{x\sigma}|0\rangle$.
When the system is with 
homogeneous distribution of holes like the holon-condensed state,
$\bar{\phi}_{x}\phi_{x}-\rho \sim 0 $ and 
$\Delta \sim 0$.
On the other hand, when the system enters into 
an inhomogeneous hole-localized state, 
$\bar{\phi}_{x}\phi_{x}-\rho \sim -\rho $ for  hole-free sites
and $\bar{\phi}_{x}\phi_{x}-\rho \sim 1-\rho $ for sites  occupied by
holes, and  $\Delta$ develops from zero.

Similarly, $\Delta_{\rm c}$ measures correlation of fluctuations of
hole densities at the NN sites.
When the deviation of hole densities from their average,
$\bar{\phi}_{x}\phi_{x}-\rho$,
at NN sites 
have similar values (same signs), $\Delta_{\rm c}$ develops 
a value of O(1). 
A hole-rich region contributes amount of $\sim (1-\rho)^2$ to $\Delta_{\rm c}$ 
and a hole-free region contributes $\sim \rho^2$.
If $\bar{\phi}_x\phi_x$ takes finite but random values 
site by site, $\Delta_{\rm c} \sim 0$ although $\Delta \sim O(1)$.  
In the phase-separated state,  both $\Delta$ and
$\Delta_{\rm c}$ should take values of $O(1)$.

\begin{figure}[t]
 \begin{center}
\includegraphics[width=7.5cm]{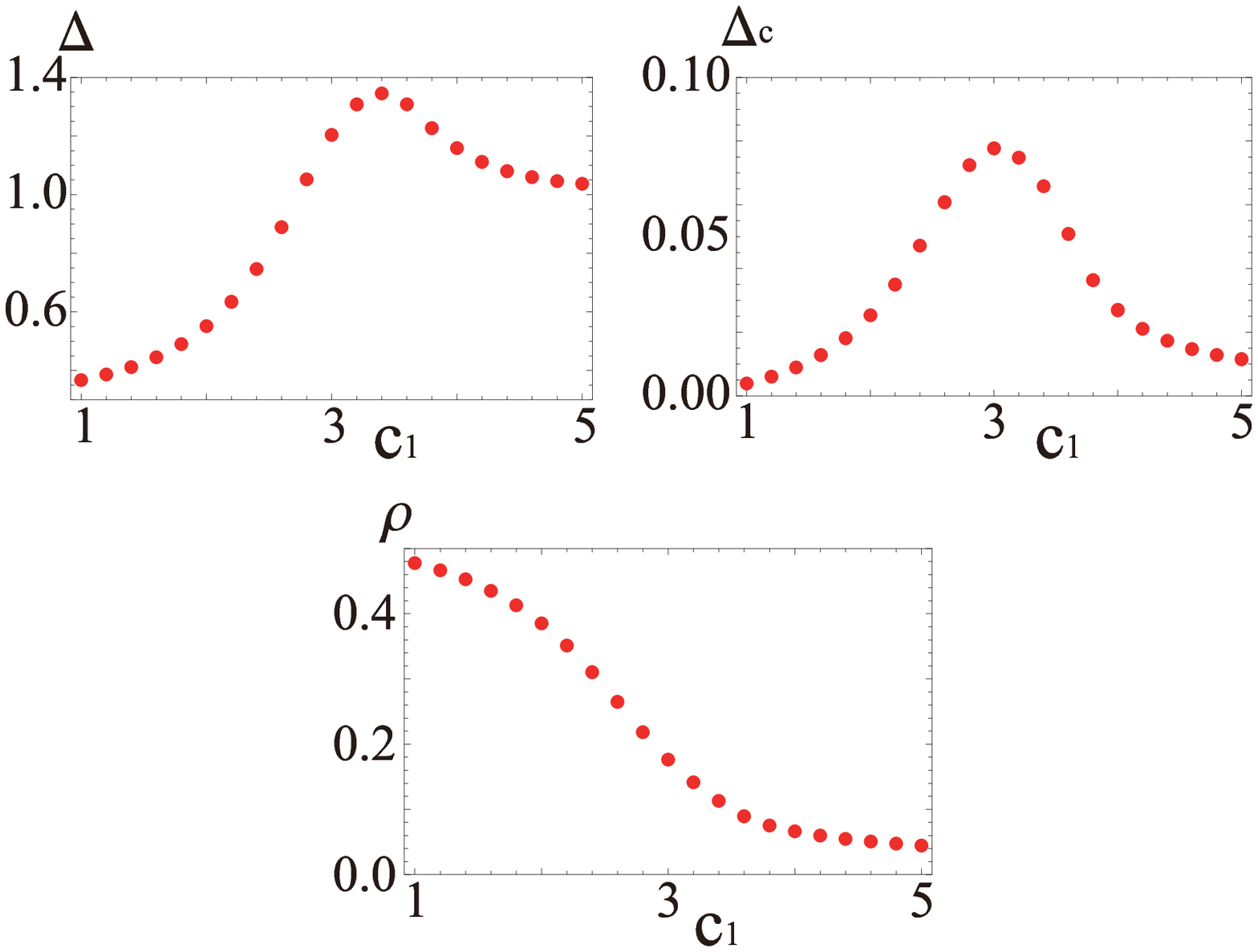}
\end{center}
\vspace{-0.3cm}
 \caption{$\Delta$ and $\Delta_{\rm c}$ 
 of (\ref{psindex})  in Model I 
 for $c_3=1.0$ and  $\mu_c=0.0$ vs $c_1$.
They have peaks around the  phase transition point
 between the PM and AF phases, $c_1\simeq 3.6$.
}
\label{PS1}
\end{figure}

\vspace{1cm}
\begin{figure}[b]
\begin{center}
\includegraphics[width=7.5cm]{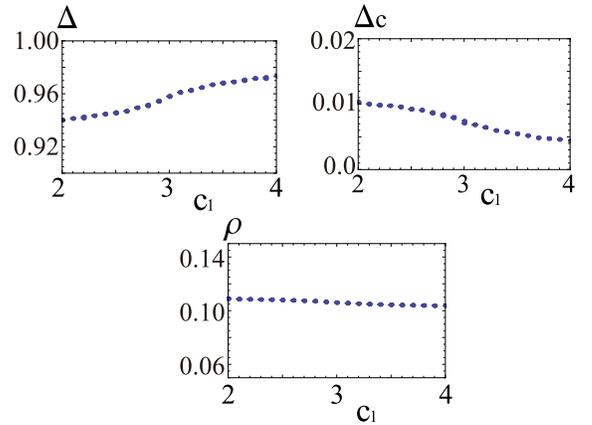}
\end{center}
\vspace{-0.3cm}
\caption{$\Delta$ and $\Delta_{\rm c}$ in Model II
for $c_3=5.0$, $\mu_c=10.0$ vs $c_1$.
The transition between the PM and AF phases  takes place at $c_1\simeq 2.9$.
}
\label{PS1_I}
\end{figure}

In Fig.\ref{PS1}, we first show $\Delta$, $\Delta_{\rm c}$ and $\rho$ 
of Model I vs $c_1$ for $c_3=1.0$ and $\mu_c=0.0$.
A phase transition from the PM to AF phases takes place at $c_1\simeq 3.6$.
Both $\Delta$ and $\Delta_{\rm c}$ 
exhibit peaks around the transition point, though the magnitude
of $\Delta_{\rm c}$ itself is very small for all $c_1$'s.
This behavior around the transition point can be understood as a 
result of formation of AF domains of finite sizes in the {\it PM phase} on
approaching to the AF phase boundary.
In the AF phase, $\Delta$ has a fairly large value, which
indicates that localization of holes takes place there.
However, the small $\Delta_{\rm c}$ means that the PS does not
take place in both AF and PM phases.
$\rho$ decreases rapidly in the AF region because
holes break the AF bonds around them as explained before, costing
higher energy in the AF phase than in the PM phase.

\begin{figure}[t]
 \begin{center}
 \includegraphics[width=7.5cm]{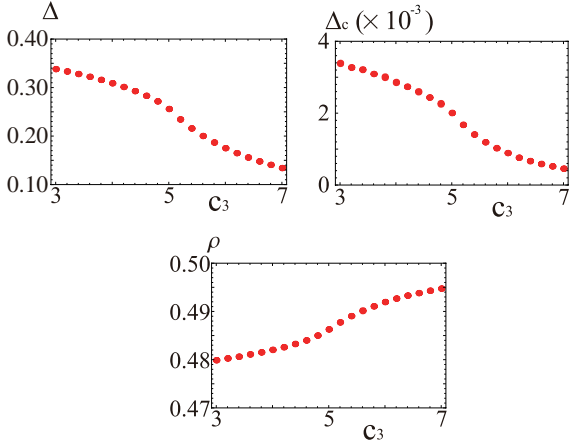}
 \end{center}
 \caption{$\Delta$ and $\Delta_{\rm c}$ in Model I
  vs $c_3$  for $c_1=1.0$, $\mu_c=0.0$.
 The PM$\leftrightarrow$FM+SF phase transition takes place 
 at $c_3\simeq 5.5$.
}
\label{PS2}
\end{figure}

\begin{figure}[b]
\begin{center}
\includegraphics[width=7.5cm]{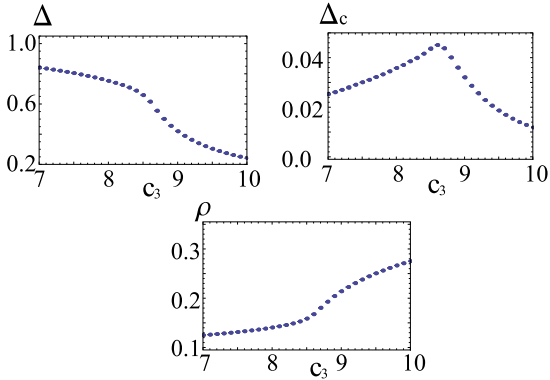}
\end{center}
\caption{$\Delta$ and $\Delta_{\rm c}$ in Model II 
vs $c_3$ for $c_1=1.0$, $\mu_c=10.0$.
The PM$\leftrightarrow$FM+SF phase transition
takes place at $c_3\simeq 8.5$.
}
\label{PS2_I}
\end{figure}

The above results should be compared with results in Model II.
See  Fig.\ref{PS1_I} for  Model II with $c_3=5.0, \ \mu_c=10.0$,
in which the PM$\leftrightarrow$AF transition takes place and 
the hole density $\rho$ takes almost the same value with 
that in the AF phase of Model I.
However, $\Delta$ and $\Delta_{\rm c}$ are almost
constant while $c_1$ is varied across  the phase transition
from the PM to AF phases.
In Model II, the spin dynamics does not influence the hole 
dynamics for sufficiently small $\delta$.
Obviously, the PS does not takes place in Model II as in 
Model I. 

Let us next see the results 
around the phase transition from the PM to FM+SF phases.
In Fig.\ref{PS2}, we show the results of Model I
for $c_1=1.0$, $\mu_c=0.0$. 
The hole density $\delta$ is fairly large, and both 
$\Delta$ and $\Delta_{\rm c}$ are decreasing functions of $c_3$.
This result is expected, as 
 holes in the SF phase are distributed homogeneously 
 through the whole system.
The corresponding results in the Model II for $c_1=1.0$ and $\mu_c=10.0$
are given in Fig.\ref{PS2_I}.
Though the above parameters have different values from those 
in Fig.\ref{PS2}, the same PM$\leftrightarrow$FM+SF phase transition 
takes place at $c_3\simeq 8.5$.
$\Delta$ exhibits similar behavior to that of the Model I,
whereas $\Delta_{\rm c}$ has a small peak at the phase transition point.

\begin{figure}[t]
 \begin{center}
\includegraphics[width=7.5cm]{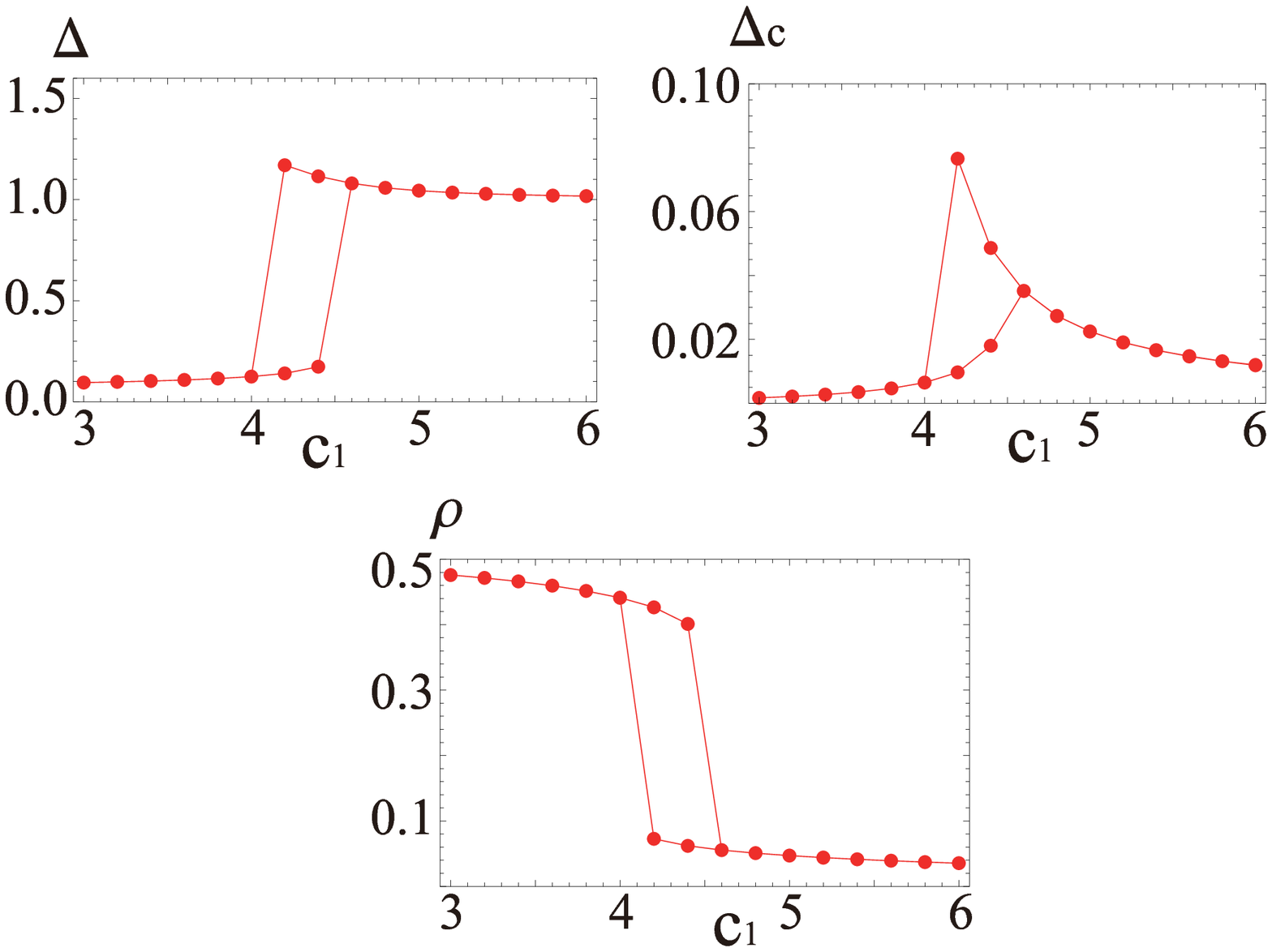}
 \end{center}
 \caption{$\Delta$ and $\Delta_{\rm c}$ in Model I
 vs $c_1$ for $c_3=10.0$, $\mu_c=0.0$. 
 The FM+SF $\leftrightarrow$AF
 phase transition take a place at $c_1\simeq 4.2$.
}
\label{PS3}
\end{figure}

\begin{figure}[b]
\begin{center}
\includegraphics[width=7.5cm]{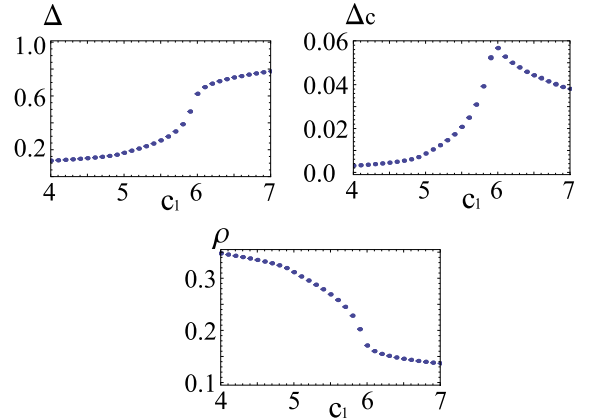}
\end{center}
\caption{$\Delta$ and $\Delta_{\rm c}$ in Model II
vs $c_1$ for $c_3=16.0$, $\mu_c=10.0$. 
The FM+SF$\leftrightarrow$AF+SF and AF+SF$\leftrightarrow$AF
phase transitions  
take place at
$c_1\simeq 5.0$ and $c_1 \simeq 5.9$, respectively.
See Fig.\ref{fig:chcs}.}
\label{PS3_I}
\end{figure}

Finally, we show $\Delta$ and $\Delta_{\rm c}$ around the
FM+SF$\leftrightarrow$AF phase transition. See  Fig.\ref{PS3}
for Model I with $c_3=10.0$, $\mu_c=0.0$.
Both quantities exhibit hysteresis loops around the transition
point $c_1\simeq 4.2$, because the phase transition is of
first order.
In the AF phase, holes are strongly localized, whereas
they have a homogeneous distribution in the FM+SF phase.
Corresponding numerical results for the Model II 
with $c_3=16.0$, $\mu_c=10.0$ are given in Fig.\ref{PS3_I}.
There exist no anomalous behavior at the transition 
FM+SF$\leftrightarrow$AF+SF at $c_1\simeq 5.0$,
because  the SF takes place in both phases.
On the other hand, $\Delta$ develops in the AF phase
as the SF disappears there.

To close this section, we compare our results
with other works.
Our conclusion that there are no PS states
is compatible with 
the results of Ref.\cite{derivation}
for the isotropic 2D model at low $T$'s,
although our system is 3D at finite $T$'s.
As mentioned in Sect.1, 
Boninsegni and  Prokof'ev\cite{btj} 
studied  the 2D bosonic $t-J$ model with anisotropic
spin coupling ($J_{x,y} = \alpha J_z, \alpha < 1)$ with 
$0 < J/t < 1$ at low $T$'s,
and found PS for low hole concentrations, $\delta < \delta_c(\alpha)$, 
where
$\delta_c(\alpha)$ is the critical density of holes
and $\delta_c(\alpha) \rightarrow 0 $ as $\alpha \rightarrow 1$.
This implies that anisotropy is essential for
PS; the stability of spins in $z$-direction  via the NN spin coupling
favors to form spin-rich regions.

\section{Conclusion and Discussion}

In the present paper, we proposed the bosonic $t$-$J$ model of
hard-core atoms and studied its physical properties by 
means of the MC simulations.
In Model II, the effects of the projection operator 
$P_x=1-\phi^\dagger_x\phi_x$ are ignored and, as a result,
there appear four phases, AF, PM, PM+SF, and AF+SF phases.
In Model I, we mostly studied the bond-breaking effects by holes and 
found that the AF+SF does not appear in the phase diagram.
We also investigated if the PS takes place and found that,
in both models, it does not. 

We argued that Model II describes effectively 
a doped AF Heisenberg model including
strong AF correlations besides the NN AF couplings,
such as {\it the next-NN spin couplings}.
Thus the important result of Model II,
appearance of the AF+SF phase, suggests that 
such a coexisting phase of AF order and SF
may be generated by inclusion of 
certain AF-enhancing couplings such as 
the next-NN coupling beyond the NN ones.
  
For the fermionic $t$-$J$ model, whether the AF+SF phase (or 
rather the coexisting phase of AF order and superconductivity) 
exists or not is an open question.
However a AF+superconducting phase has been observed by recent experiments
with uniform high-quality samples of high-$T$ cuprates\cite{mukuda}.
Therefore we expect that a similar coexisting phase of AF+SF appears in
cold-atom systems in an optical lattice with strong on-site
repulsion as long as certain AF-enhancing couplings such as 
next-NN coupling are involved.
 
Another interesting subject is to consider the effect of 
anisotropy of pseudo-spin coupling ($\alpha < 1$).
Although we studied the isotropic case ($\alpha = 1$) in the
present paper as the canonical case, the realistic cold-atom systems
may have anisotropy. 
This is because the difference in two species of bosons
such as their hopping amplitudes, etc., are reflected to this anisotropy
in the resulting $t$-$J$ model (recall the relation $J=4t^2/U$).

At the beginning of Sect.4, we mentioned the possible 
superconductivity via formation and condensation
of NN hole pairs in Model I.
Because we found  no separated phase transition into superconducting phase
in Sect.4, it is plausible that the superconducting state of hole pairs
is generated simultaneously with the SF state of single-boson BEC.
To examine this point further, it is necessary 
to study the correlation function of these pairs in details.
When the anisotropy is included, such correlation may have different 
behaviors in different pairing channels.
These topics (the next-NN coupling, anisotropy, and the hole-pairing)
are under study. 
We hope to report on them in future publications.

Finally, studies on the bosonic $t$-$J$ model in the present 
and previous papers suggest
for the fermionic $t$-$J$ model that, at finite hole concentrations,
coherent holon hopping is realized. In the slave-fermion approach, 
this implies that
a small Fermi surface of fermionic holons
is generated in under-doped region. In the further study of the bosonic
$t$-$J$ model, we expect to draw further interesting
suggestions for the fermionic $t$-$J$ model and vice versa.\\


\bigskip
\begin{center}
{\bf Acknowledgment} \\
\end{center}
We thank Dr. Kenichi Kasamatsu for useful comments.
This work was partially supported by Grant-in-Aid
for Scientific Research from Japan Society for the 
Promotion of Science under Grant No.20540264.\\

\appendix
\renewcommand{\theequation}{\Alph{section}.\arabic{equation}}
\section{CP$^1$ Representation of a Hard-core Boson}

In this Appendix we derive the representation 
(\ref{phiw}, \ref{phiw2}) of a HCB operator
in terms of a CP$^1$ spin operator $w_{x\eta}$.

Let us introduce an $s=1/2$ SU(2) (pseudo) spin operator
$\vec{T}_x$ by using $w_{x\eta}$ as
\be
\vec{T}_x &=& \frac{1}{2}\bar{w}_x \vec{\sigma} w_x,
\ee
where $\vec{\sigma}$ are $2\times 2$ Pauli matrices.
Explicitly,
\be
T_{x1} &=& \frac{1}{2}(\bar{w}_{x1}x_{x2}+\bar{w}_{x2}x_{x1}),\nn
T_{x2} &=& -\frac{i}{2}(\bar{w}_{x1}x_{x2}-\bar{w}_{x2}x_{x1}),\nn
T_{x3} &=& \frac{1}{2}(\bar{w}_{x1}x_{x1}-\bar{w}_{x2}x_{x2}).
\ee

On the other hand,
the rising operator $T^+_x$ and the lowering operator
$T^-_x$ defined by
\be
T^{\pm}_x \equiv T_{1x}\pm i T_{2x},
\ee
is expressed by the HCB $\phi_x$ as
\be
T^+_x = \phi^\dag_x,\ T^-_x = \phi_x.
\ee
Then we have
\be 
\phi_x = w_{x2}^\dag w_{x1},
\ee
which is (\ref{phiw}, \ref{phiw2}).
The correspondence between the two sets of states are as follows;
\be
T_3|+\ra &=& \frac{1}{2}|+\ra,\ T_3|-\ra = -\frac{1}{2}|-\ra,\nn
\phi|0\ra &\equiv&0,\ |-\ra= |0\ra,\ |+\ra= \phi^\dag|0\ra, \nn
w_{\eta}|v\ra&\equiv&0,\ |-\ra= w^\dag_{2}|v\ra,\ |+\ra= w^\dag_{1}|v\ra.
\ee\\



\begin{references}
\bibitem{B0} 
We write the HCB operator as $B_{x\sigma}$ instead of
$\hat{B}_{x\sigma}$ for simplicity.
\bibitem{hcb}
The hard-core nature may be implemented
by starting from the usual bosons without hard-core nature,
adding to the Hamiltonian (\ref{Hhub}) the term
$V\sum_{x\sigma}\hat{n}_{x\sigma}(\hat{n}_{x\sigma}-1)$,
 and taking the limit $V \to \infty$. This term allows only
for the states $\hat{n}_{x\sigma}=0,1$ and excludes 
the states $\hat{n}_{x\sigma} > 1$.
\bibitem{acr}
$[A,B]_+ \equiv AB + BA$ is the anticommutation relation.
\bibitem{3sites}
There appears another term
$\sum_x\sum_{\mu, \nu}\epsilon_{\sigma\sigma'}B_{x\sigma}^\dag B_{x+\mu,\sigma'}
\times \epsilon_{\tau\tau'}B_{x+\mu,\tau}^\dag B_{x+\nu\tau'}$.
However, their coefficient has an extra factor $\propto \delta$
and can be negligible for small hole concentrations.
\bibitem{doubleoccupancy}
This restriction to the physical space is exactly the same as in the 
fermionic $t$-$J$ model.
\bibitem{2s}
For general $s$, the constraint (\ref{const1}) becomes
$(2s)^{-1}\sum_{\sigma=1}^2 a_{x\sigma}^\dag a_{x\sigma} +
\phi_x^\dag\phi_x =1$. There are $2s+1$ possible physical states for each $x$.
For example, 
the  states for $s=1$ are described by
$(z^\dag_{x\uparrow}z^\dag_{x\uparrow},\ 
z^\dag_{x\uparrow}z^\dag_{x\downarrow},\ z^\dag_{x\downarrow}
z^\dag_{x\downarrow})|0\ra$.
\bibitem{B} Again we used the same letter $B_{x\sigma}$
as the operator (\ref{B}) because no confusions arise.
\bibitem{FN1}As we are considering a quantum system, 
a state at each site is generally given by a superposition of 
the holon state and the spinon state.
The quantum state that is totally occupied by a holon 
is a very specific state.
\bibitem{2d3d}
For the 2D system, the number of broken AF bonds are
seven for a NN holon pairs and eight for a more separated pair.
\end{references}
\end{document}